\documentclass[12pt]{article}
\pdfoutput=1 
\usepackage{slashed}
\usepackage{color, verbatim}
\usepackage{latexsym}
\usepackage{amssymb}
\usepackage{amsmath}
\usepackage{graphicx}
\graphicspath{{figs/}}
\usepackage{arydshln}
\usepackage[dvipsnames]{xcolor}
\usepackage{adjustbox}
\usepackage{easybmat}
\usepackage{bbm}
\usepackage{cite}
\usepackage{slashed}
\usepackage{xcolor}
\usepackage{bm}
\usepackage{adjustbox}
\usepackage{booktabs}
\usepackage{multirow} 
\usepackage{rotating}
\usepackage{mathtools}
\usepackage[inline]{enumitem}
\usepackage{lscape}
\usepackage{booktabs}
\usepackage{subcaption}
\usepackage[normalem]{ulem}
\usepackage{subcaption}
\usepackage{multirow}
\usepackage{makecell}
\usepackage{booktabs}
\usepackage[colorinlistoftodos]{todonotes}
\usepackage{hyperref}
\usepackage{enumitem}
\DeclareMathOperator{\tr}{Tr}
\DeclareMathOperator{\llog}{Log}
\DeclareMathOperator{\tev}{TeV}
\renewcommand{\theequation}{\thesection.\arabic{equation}}
\makeatletter
\@addtoreset{equation}{section}
\g@addto@macro\bfseries{\boldmath}
\newcommand\Label[1]{&\refstepcounter{equation}(\theequation)\ltx@label{#1}&}
\makeatother
\allowdisplaybreaks

\begin{document}
	%
	\thispagestyle{empty}
	\begin{flushright}
	\end{flushright}
	\vspace{0.8cm}
	
	\begin{center}
		{\Large\sc A bridge to new physics: proposing new -- and reviving old -- explanations of $a_\mu$
		}
		\vspace{0.8cm}
		
		\textbf{
			Guilherme Guedes$^{\,a,b}$ and Pablo Olgoso$^{\,a}$ }\\
		\vspace{1.cm}
		{\em {$^a$ CAFPE and Departamento de F\'isica Te\'orica y del Cosmos,
				Universidad de Granada, Campus de Fuentenueva, E--18071 Granada, Spain}}\\[0.5cm]
		{\em {$^b$ Laborat\'orio de Instrumenta\c cao e F\'isica Experimental de Part\'iculas, Departamento de
				F\'isica da Universidade do Minho, Campus de Gualtar, 4710-057 Braga, Portugal}}\\[0.2cm]

		\vspace{0.5cm}
	\end{center}
	\begin{abstract}
		The $4.2\,\sigma$ tension in the combined measurement of the anomalous magnetic moment of the muon, $a_\mu$, and the Standard Model prediction strongly suggests the existence of beyond the Standard Model physics. Following the Standard Model Effective Field Theory approach, we study a particular topology, the \emph{bridge} diagram, which gives a chirally enhanced contribution to $a_\mu$. We classify all possible 2- and 3-field SM extensions that can generate this contribution and present the full $a_\mu$ result for them. Within our approach, we find that several 2-field fermion-scalar extensions which had been previously discarded in the literature -- when only the Yukawa-suppressed contribution was considered -- can actually be viable models to explain the observed anomaly. Furthermore, the 3-field extensions which generate the bridge diagram represent a new class of models to account for $a_\mu$. We explore a particular 3-field extension which, beyond explaining $a_\mu$, can also account for the neutral $B$-meson anomalies and the Cabibbo angle anomaly. We present the full one-loop matching for this model and a one-loop phenomenological study.
	\end{abstract}
	
	\newpage
	
	\tableofcontents
	
	\section{Introduction}
	The recent measurement of the anomalous magnetic moment of the muon, $a_\mu$, by Fermilab Muon g-2 Experiment \cite{Muong-2:2021ojo}, has sparked immense interest in the physics community since, together with the previous measurement by Brookhaven National Laboratory \cite{Muong-2:2006rrc}, they present a combined $4.2\sigma$ tension with the Standard Model (SM) result \cite{Aoyama:2020ynm,davier:2017zfy,keshavarzi:2018mgv,colangelo:2018mtw,hoferichter:2019mqg,davier:2019can,keshavarzi:2019abf,kurz:2014wya,chakraborty:2017tqp,borsanyi:2017zdw,blum:2018mom,giusti:2019xct,shintani:2019wai,FermilabLattice:2019ugu,gerardin:2019rua,Aubin:2019usy,giusti:2019hkz,melnikov:2003xd,masjuan:2017tvw,Colangelo:2017fiz,hoferichter:2018kwz,gerardin:2019vio,bijnens:2019ghy,colangelo:2019uex,pauk:2014rta, danilkin:2016hnh,jegerlehner:2017gek,knecht:2018sci,eichmann:2019bqf,roig:2019reh,colangelo:2014qya,Blum:2019ugy,    aoyama:2012wk,Aoyama:2019ryr,czarnecki:2002nt,gnendiger:2013pva},
	\begin{equation}
		a^{\mathrm{EXP}}_\mu - a^{\mathrm{SM}}_\mu = (251 \pm 59) \times 10^{-11},
	\end{equation}
	which strongly suggests the existence of new physics (NP) contributions to this quantity. This value does not take into account the latest lattice results from the hadronic vacuum polarization SM contributions, presented by the BMW collaboration \cite{Borsanyi:2020mff}, which seem to soften the tension in the $a_\mu$ measurement; however, since it introduces another tension in $e^+e^- \rightarrow$ hadrons cross-section \cite{Keshavarzi:2020bfy,Crivellin:2020zul,Malaescu:2020zuc,Colangelo:2020lcg}, we will neglect it for now.
	
	From the lens of the Standard Model Effective Field Theory (SMEFT), the new contribution to $a_\mu$ (at tree-level) is given by
	\begin{equation}
		\label{eq:amu}
		\Delta a_\mu = a_\mu^{\mathrm{NP}} - a_\mu^{\mathrm{SM}} = \frac{4 m_\mu v}{\sqrt{2} e} \bigg( \mathrm{Re}[\alpha_{eB}^{2,2}] c_W - \mathrm{Re}[\alpha_{eW}^{2,2}] s_W \bigg) \equiv \frac{4 m_\mu v}{\sqrt{2} e} \alpha^{2,2}_{e\gamma}  \,,
	\end{equation}
	where $m_\mu$ is the mass of the muon, $v$ the vacuum expectation value (VEV) of the Higgs, $e$ the electric charge, $c_W$ ($s_W$) the co-sine (sine) of the Weinberg angle and $\alpha^{i,j}_{eB}$ and $\alpha^{i,j}_{eW}$ the coefficients of the dipole operators,
	\begin{align}
		\mathcal{O}^{i,j}_{eB} &= (\overline{\ell}^i \sigma^{\mu\nu} e^j) H B_{\mu\nu} + \mathrm{h.c.}, \\
		\mathcal{O}^{i,j}_{eW} &=(\overline{\ell}^i \sigma^{\mu\nu} e^j)\sigma^I H W^I_{\mu\nu} + \mathrm{h.c.}\,,
	\end{align}
	where $i$ and $j$ correspond to flavour indices.
	Below the electroweak scale, in the low-energy effective field theory (LEFT), the tree-level contribution to $a_\mu$ is given by the photon dipole operator, whose Wilson coefficient is $L_{e\gamma}= \frac{v}{\sqrt{2}} \alpha_{e\gamma}$. Results in this paper will be quoted as contributions to $\alpha_{e\gamma}$.
	
	In spite of contributing at tree-level to $a_\mu$, the SMEFT dipole operators are generated only at loop level by weakly-coupled UV completions \cite{Arzt:1994gp,deBlas:2017xtg,Craig:2019wmo} and, as such, a complete one-loop analysis of the contributions to $a_\mu$ needs to include the effects of the one-loop running of tree-level generated operators and one-loop finite contributions at low-energy. In Ref. \cite{Aebischer:2021uvt} it was shown that, besides the mentioned dipole operators, the only remaining relevant contribution to $a_\mu$ arises as a one-loop effect from $\mathcal{O}_{\ell e q u}^{(3)}$ for an EFT with a cut-off of $\Lambda = 10\; \mathrm{TeV}$. Being already a one-loop contribution, one only needs to consider the tree-level generation of this operator, which can only occur by integrating out the scalar leptoquarks $S_1$ and $S_2$ \cite{deBlas:2017xtg}; since these tree-level contributions are already well known \cite{deBlas:2017xtg,Aebischer:2021uvt}, for the remainder of this paper we will focus solely on the contribution to $a_\mu$ arising from the one-loop generation of the dipole operators.

	Several completions to the SM (with a different number of new fields) have been proposed in the literature to generate the dipole operators -- see Ref. \cite{Athron:2021iuf} (and references therein) for a comprehensive review of the status of such models. Interest has been given to chirally enhanced solutions, that is, solutions in which the new contribution to $a_\mu$ is not suppressed by the muon Yukawa coupling, because they can explain the observed $\Delta a_\mu$ with masses for the new heavy particles typically large enough to avoid current experimental constraints. In particular, Refs. \cite{Calibbi:2018rzv,Calibbi:2019bay,Arnan:2019uhr,Crivellin:2021rbq,Allwicher:2021jkr,Arcadi:2021cwg} studied chirally enhanced contributions for a wide range of 3-field extensions of the SM.
	
	In this work, we focus on the chirally enhanced contribution to $a_\mu$ produced by the topology of Fig. \ref{fig:ffs}, which we will refer to as \emph{bridge} hereafter. We will study the possible UV completions which can generate this diagram. Note that the particle connecting the loop and 2 external states (the bridge) and one of the particles in the loop must be heavy, but the other one can be heavy or light, and as such, either 2- or 3-field extensions can generate this type of topology.

	While the bridge diagram has been studied in some particular cases of 2-field extensions with 2 vector-like leptons \cite{Arkani-Hamed:2021xlp,Rose:2022njd}, we perform a complete classification of all possible 2 and 3 fields extensions of the SM which can produce this topology. These extensions can sometimes be accompanied by a box diagram\footnote{In the LEFT, when the Higgs takes a  VEV, this actually corresponds to the triangle diagram which is usually considered in the literature.} which one must take into account as well when computing $a_\mu$. 
	
	Due to our approach in the SMEFT, we consider contributions from several 2-field extensions which have been overlooked in the literature, where only the lepton Yukawa-suppressed contributions to $a_\mu$ were considered excluding these models as explanations of the anomaly either by direct searches or due to the negative sign in their contribution. By considering the bridge contribution from these 2-field models, we are in fact restoring them as possible explanations for the observed anomaly in $a_\mu$.
	
	The 3-field extensions with chirally enhanced contributions considered in Refs. \cite{Calibbi:2019bay,Arnan:2019uhr,Crivellin:2021rbq,Allwicher:2021jkr,Arcadi:2021cwg,Calibbi:2018rzv} generated the dipole operators through the box diagram, shown in Fig. \ref{fig:trianglesff}. The representations of the heavy fields which fit into that topology are different from the ones which can generate the bridge diagram, and, as such, in this work we present a completely new class of 3-field extensions that can account for the observed $\Delta a_\mu$.
	
	Given the added flexibility of considering 3 new degrees of freedom, we consider a particular set of representations for the 3 fields in this class in which we can also address the neutral flavour anomalies $R_K$ and $R_{K^*}$, as well as the Cabibbo angle anomaly. This model is composed of an $SU(2)$ triplet leptoquark, $S_3$, a triplet vector-like lepton, $\Sigma$, and a triplet vector-like quark, $\Psi_Q$. We present here the full one-loop matching conditions for this model and study its one-loop phenomenology.

	The article is organized as follows. In Section 2 we explore the technicalities of the computation of the matching conditions for the $a_\mu$ contribution. In Section 3 we provide generic results for the bridge topology for the different combinations of heavy scalar and heavy fermion propagators that can generate it. Section 4 is devoted to presenting the results of $\Delta a_\mu$ for all 2-field extensions which generate the bridge topology, featuring novel results for a particular set of completions. In Section 5 the same is done for 3-field extensions, introducing a new class of models which can explain $\Delta a_\mu$. Phenomenological considerations which can be applied to all of these models are discussed in Section 6. In Section 7 we perform the complete one-loop matching for a particular 3-field extension, in which the neutral flavour anomalies and the Cabibbo angle anomaly can also be addressed.

	\section{Computation of $a_\mu$}
	\label{sec:computationamu}

	In this work we consider fermion and scalar\footnote{Scalars can be replaced by heavy vectors with the same quantum numbers, but we do not present results for those cases.} 2- and 3-field extensions of the SM which generate the bridge topology of Fig. \ref{fig:ffs} -- with the $W$ or $B$ gauge bosons attached on either of the internal propagators\footnote{Note that in practice one can directly calculate the diagram with the photon insertion and consider the appropriate electric charge; however, to keep the language coherent within the SMEFT picture, we will refer to diagrams with both $W$ and $B$ bosons.}. Throughout all computations we neglect contributions suppressed by lepton Yukawa couplings or by the Higgs mass (terms of the form $m_\phi/M$).
	
	Some of the extensions we consider can also contribute to $a_\mu$ through other diagrams, namely the usual box diagrams shown in Figs. \ref{fig:bridgescalarw2fermionloop} and \ref{fig:trianglesff}. Therefore, for completeness, we present in Appendix A the contribution to $a_\mu$ for general representations of heavy fields arising from box diagrams. With this and the general results from the bridge topology, which are presented in the next section, one can in principle calculate the full contribution to $a_\mu$ for arbitrary UV extensions of the SM. Furthermore, note that depending on the particular representation, some 3-field extensions can also generate diagrams with only 2 or 1 heavy propagators, which have different kinematic factors and must also be considered in the calculation of the full $a_\mu$ contribution\footnote{An example of this are 3-field extensions in which a heavy Higgs is considered, where one must also consider the diagrams generated by substituting it by the SM Higgs. Note also that the heavy Higgs by itself can generate a contribution to $a_\mu$.}.  When we present results for a particular SM extension (in Sections 4 and 5), we consider all possible contributions to $a_\mu$. 
	
	To obtain the one-loop matching conditions of the dipole operators, 
	we compute the 4 point amplitude between $\overline{\ell}_L$, $ e_R$, $ \phi$, and $B/W$,
	with all momenta incoming, in the full theory and in the SMEFT. For simplicity, we take the momentum of the Higgs, $p_\phi$, to zero. The kinematic structure which uniquely defines the dipole operators is $\slashed{q} \slashed{\epsilon}$, where $q$ is the gauge boson momentum and $\epsilon$ its polarization vector. This structure can be traded on-shell by $\epsilon\cdot p_e$, since:
	\begin{align}
		\label{eq:relation}
		\overline{v_{\ell}}\,\slashed{q} \slashed{\epsilon} \,u_{e} = - \overline{v_{\ell}}\,(\slashed{p}_\ell + \slashed{p}_e) \slashed{\epsilon}\,u_{e} = - 2 \epsilon\cdot p_e \,\overline{v_{\ell}}\,u_{e}\;,
	\end{align}
	where $p_{e(\ell)}$ is the momentum of the right-handed (left-handed) electron, and $v_{\ell},u_e$ are the corresponding external spinors.
	In the second equality we considered the on-shell conditions $\overline{v_{\ell}}\,\slashed{p}_\ell  = 0$ and $\slashed{p}_e\,u_e=0$ (when applied to the external fields for massless fermions). Therefore, $\epsilon\cdot p_e$ ends up being the only relevant kinematic structure for the matching calculation.

	For this specific case, performing the matching on-shell is particularly efficient since no other connected diagrams can arise (without inserting lepton Yukawas).
	One could think of attaching the gauge bosons to the external legs of the diagrams instead of in the internal propagators. However, when the gauge boson is attached to the fermions, the diagram will either result in a contribution proportional to $\slashed{p}_\ell\slashed{\epsilon}$ or $ \slashed{\epsilon}\, \slashed{p}_e $, or,  when the photon couples to the Higgs, proportional to $q\cdot\epsilon= 0$  or $p_\phi\cdot \epsilon=0$.
	
	The same happens when the gauge boson is attached to the fermionic bridge, where one can find that all contributions are proportional either to $\slashed{p}_\ell \slashed{\epsilon}$ or $\slashed{\epsilon}\,\slashed{p}_e$, being therefore zero in light of the arguments presented above. Consequently, we only compute contributions coming from insertions of gauge bosons in the particles in the loop. For the same reasons, a mass insertion in the bridge propagator is needed, fixing the chirality of the coupling between the two (three) heavy fields.
	
	All results presented in Sections 3 and 4 were cross-checked with {\tt matchmakereft} \cite{Carmona:2021xtq}~\footnote{When possible we also cross-checked our results against those in the literature.}. This tool calculates the one-loop matching conditions for UV extensions of the SM diagramatically and off-shell. Therefore, results in this case are given in terms of operators in a Green's basis, which must then be reduced to the Warsaw basis. When using {\tt matchmakereft} one must then take
	\begin{align}
		\alpha_{eB} &= \alpha^G_{eB}-\frac{g_1}{8} \beta^G_{eHD2} + \frac{g_1}{8} \beta^G_{eHD4} - \frac{g_1}{2} \beta^G_{eHD3}\,,\\
		\alpha_{eW} &= \alpha^G_{eW} - \frac{g_2}{8} \beta^G_{eHD2} + \frac{g_2}{8} \beta^G_{eHD4}\,,
	\end{align}
	where all the coefficients are written following the conventions stated in {\tt matchmakereft}, i.e., $\beta$ for redundant operators, and the superscript $G$ corresponds to Wilson coefficients in the Green's basis. We neglected Yukawa-suppressed contributions and the evanescent coefficients $\gamma_{eB},\gamma_{eW}$. Following from Eq. (\ref{eq:amu}), one can then write the contribution to $a_\mu$ as a function of the Wilson coefficients in the Green's basis as:
	\begin{equation}
		\Delta a_\mu= \frac{4 m_\mu v}{\sqrt{2}} \bigg( \frac{1}{g_1}\,\left(\alpha^G_{eB}\right)^{2,2}-\frac{1}{g_2}\,\left(\alpha^G_{eW}\right)^{2,2}-\frac{1}{2}\,\left(\beta^G_{eHD3}\right)^{2,2} \bigg).
	\end{equation}

	\section{General results}
	\label{sec:generic}
	The bridge topologies can be divided according to the particle that runs in the bridge, where by bridge we are referring to the internal propagator connecting the loop with 2 external particles. There can never be a SM particle in the bridge as it would give a Yukawa-suppressed contribution. The possible heavy particles in the bridges are fixed by the external SM particles to which it couples to:

	\begin{enumerate}[label=\textbf{\arabic*}.]
		\item{\textbf{Scalar bridge}}
		
		The heavy scalar must couple to the left- and right-handed muon, which fixes the quantum numbers to be the same as the SM Higgs.  All the corresponding contributions to $a_\mu$ are zero (they are always proportional to $\epsilon\cdot q$).
		
		\item{\textbf{Fermion bridge coupled to right-handed muon}}
		
		The heavy fermion must have the same quantum numbers as the SM left-handed lepton, $\Delta\sim (1,2,-1/2)$. The numbers in parenthesis denote the representations under $SU(3)_c$, $SU(2)_L$ and $U(1)_Y$, respectively.
		
		\item{\textbf{Fermion bridge coupled to left-handed muon}}
		
		The heavy fermion must either have the quantum numbers of a SM right-handed lepton, $E\sim (1,1,-1)$, or be an $SU(2)$ triplet, $\Sigma\sim(1,3,-1)$.

	\end{enumerate}

	We present the results for the contribution to $a_\mu$ from the bridge topologies for generic representations of the heavy fields for these different vector-like lepton (VLL) bridges in the next subsections.

	For concreteness, the results are always presented corresponding to one specific orientation of the internal propagators shown in the diagrams (in particular, we always avoid the presence of fermion-number violating interactions). To translate from these results to those with a flipped propagator -- which may be needed for some choices of the gauge representations of the heavy fields --, it is sufficient to add a minus sign in the contribution corresponding to a gauge boson insertion in the flipped propagator.

	\subsection{VLL doublet bridge}
\begin{figure}
		\centering
		\begin{subfigure}[b]{0.45\textwidth}            
			\includegraphics[width=\textwidth]{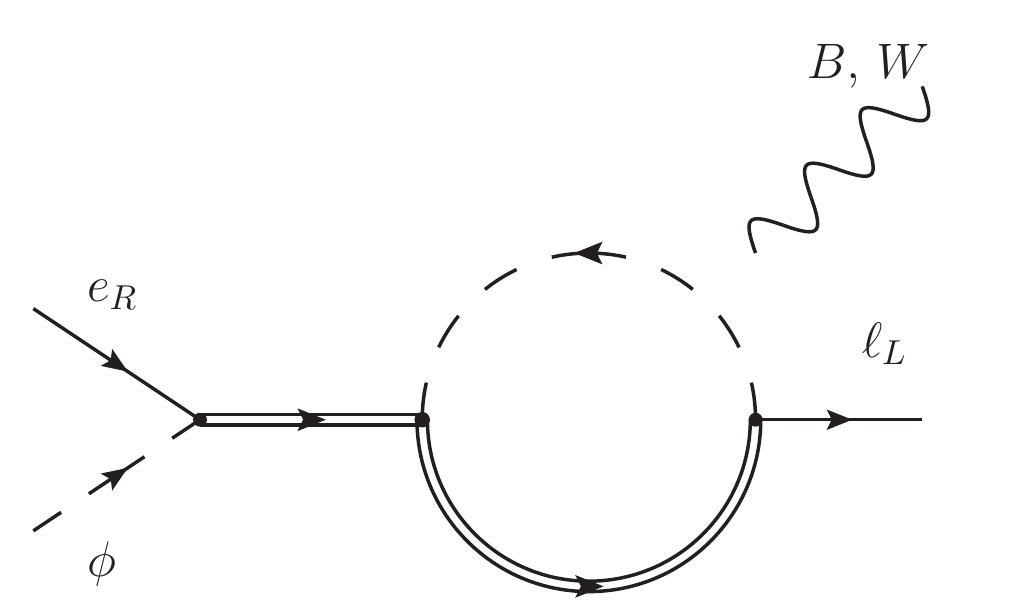}
			\caption{}
			\label{fig:ff}
		\end{subfigure}
		\begin{subfigure}[b]{0.45\textwidth}
			\includegraphics[width=\textwidth]{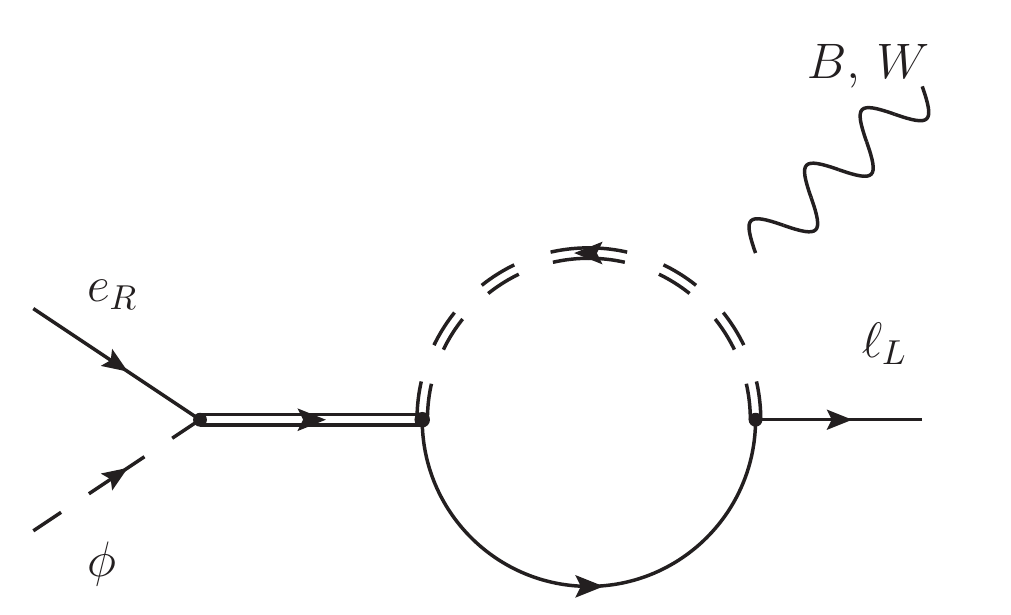} 
			\caption{}
			\label{fig:fs}
		\end{subfigure}
		\caption{\emph{Left:} Bridge topology  for the fermionic bridge with an extra heavy fermion and the SM Higgs. \emph{Right:} Bridge topology  for the fermionic bridge with an extra heavy scalar and a SM fermion. Double lines represent heavy particles whereas single lines are SM particles. The gauge boson ($B$ or $W$) is represented outside the diagram since it can be attached to any of the internal propagators.}
		\label{fig:bridgesfermion}
	\end{figure}

\begin{figure}
		\centering
		\includegraphics[width=0.5\textwidth]{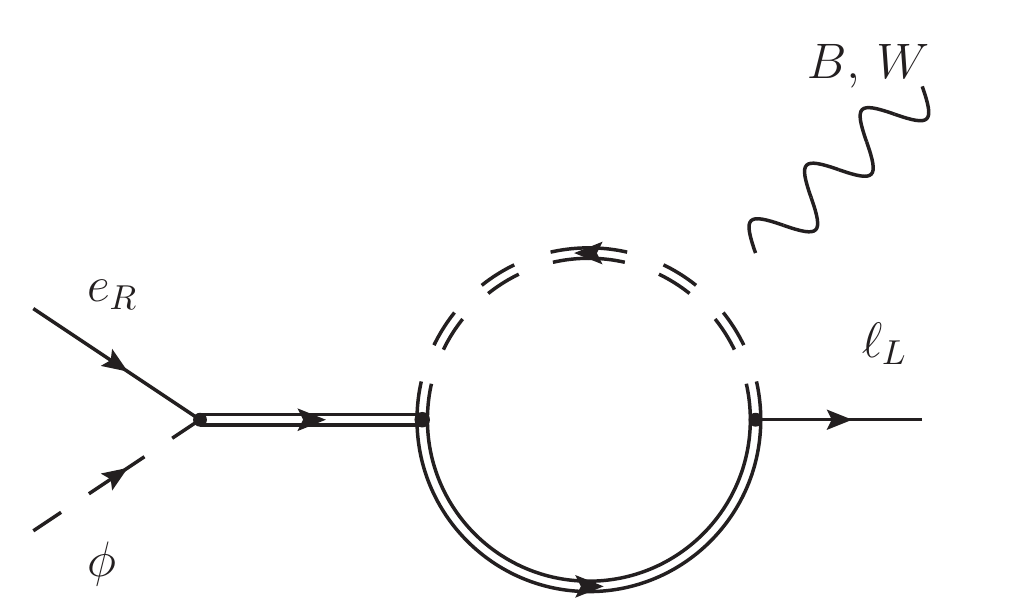}

		\caption{Bridge topology for the fermionic bridge with an extra heavy fermion and a heavy scalar. Double lines represent heavy particles, whereas single lines are SM particles. The gauge boson ($B$ or $W$) is represented outside the diagram since it can be attached to any of the internal propagators.}\label{fig:ffs}
	\end{figure}

	When the bridge particle is a heavy fermion, we are left with 3 possibilities for the combinations inside the loop: 1 extra heavy fermion, $\Psi$, and the SM Higgs, Fig. \ref{fig:ff}; 1 heavy fermion and 1 heavy scalar, $\Phi$, Fig. \ref{fig:ffs}; 1 heavy scalar and an SM fermion, Fig. \ref{fig:fs}. The latter case of Fig. \ref{fig:fs} does not give a contribution to $a_\mu$, and as such, it will be neglected in the following discussion. The reason is that, since a mass insertion in the bridge propagator is needed (as explained in Section 2), an extra mass insertion is required from the fermionic propagator in the loop in order to get the correct chirality for the external fermions.
	
	The most general Lagrangian, extending the SM with the VLL doublet, $\Delta$, that can generate a bridge-like contribution to $a_\mu$ is the following:
	\begin{align}
		\label{eq:lagdoublet}
		\mathcal{L}\supset& \, gY_\Psi\overline{\Psi}\gamma_\mu\Psi B^\mu +g_W T^{W,\Psi}_{IKI'}\overline{\Psi_I}\gamma_\mu\Psi_{I'} W^\mu_K 
		\\&-igY_\Phi B^\mu (\partial_\mu\Phi^\dagger\Phi -\Phi^\dagger\partial_\mu\Phi) -i g_W T^{W,\Phi}_{JKJ'} W^{\mu}_K (\partial_\mu\Phi^\dagger_J\Phi_{J'} -\Phi^\dagger_J\partial_\mu\Phi_{J'})
		\nonumber \\& +y_M \overline{\Delta} \,e_R \,\phi +T_{IKJ} \left(y_b^R\,  \overline{\Psi}_I \, P_R \,\Delta_K \Phi_J+ y_b^L\,   \overline{\Psi}_I \, P_L \,\Delta_K \Phi_J \right) + y_F\,T'_{KIJ}\, \overline{\ell_{L,K}}\Psi_I \Phi^{\dagger}_J+\mathrm{h.c.}\, ,\nonumber
	\end{align}
	where $\Phi$ can stand generically for a heavy scalar or the SM Higgs ($\phi$ always stands for the SM Higgs), $\Psi$ is a heavy fermion and $\ell_L$ is the left-handed SM lepton. We do not write family indices in the couplings with the SM leptons as we assume only the needed couplings to muons.
	The indices $I^{(')}$, $J^{(')}$, $K^{(')}$ correspond to the SU(2) components of the fields, $Y_{\Psi(\Phi)}$ denotes the hypercharge of $\Psi(\Phi)$, and $T^W$, $T$ and $T'$ are the Clebsch-Gordan coefficients of the fields in the corresponding interaction term. For instance, for a colourless SU(2) triplet $\Psi$ and SU(2) doublet $\Phi$, one could have $T^{W,\Psi}_{IKJ}=i\epsilon_{IKJ}$, $T^{W,\Phi}_{IKJ}=\sigma^K_{IJ}/2$, $T_{IKJ}=(\epsilon\sigma^I)_{KJ}$, $T'_{KIJ}=(\sigma^I\epsilon)_{KJ}$, with $\epsilon$ denoting the Levi-Civita tensor and $\sigma^i$ Pauli matrices.

	With this notation, and defining $T^{\gamma,\psi}_{ij}\equiv Y_{\psi} \delta_{ij}+T^{W,\psi}_{i3j}$, where $\psi$ represents any particle, we can write the generic result for $\alpha_{e\gamma}$ from the bridge topology as:
	\begin{equation}
		\label{eq:gendoublet}
		\alpha_{e\gamma}^{2,2}=\frac{iN e }{4}y_M y_Fy_b^R\sum_{IJ} T_{I2J}\left[ \gamma_\Psi T^{\gamma,\Psi}_{I'I} T'_{2JI'}  + \gamma_\Phi T^{\gamma,\Phi}_{JJ'}T'_{2IJ'} \right]\,,
	\end{equation}
	with $\gamma_{\Psi,\Phi}$ being kinematic factors which will be defined below, corresponding to the insertion of the gauge bosons on the fermion and scalar, respectively; their explicit expression depends on the number of heavy propagators. $N$ designs the dimension of the $SU(3)$ representation of $\Psi$ (and $\Phi$, by extension, when denoting a heavy scalar). $T^{\gamma,\psi}$ would be diagonal and proportional to the electric charge if the charge eigenstate basis is chosen for the $\psi$ multiplet, i.e., $T^{W,\psi}$ is diagonal. 
	
	The kinematic factors $\gamma_{\Psi,\Phi}$ defined throughout Section 3 can always be expressed as these two functions of the masses:
	\begin{align}
		f(M_A,M_B,M_C)&\equiv-\frac{iM_B}{(4\pi)^2M_A}\,\frac{M_B^4-4M_B^2M_C^2+3M_C^4+2M_C^4\llog{[M_B^2/M_C^2]}}{(M_B^2-M_C^2)^3}\,,\\
		h(M_A,M_B,M_C)&\equiv -\frac{iM_B}{(4\pi)^2 M_A}\,\frac{M_B^4-M_C^4-2M_B^2M_C^2\llog{[M_B^2/M_C^2]}}{(M_B^2-M_C^2)^3}\,.
	\end{align}
	
	For the case in which inside the loop we have another heavy fermion, $\Psi$ and the SM Higgs, Fig. \ref{fig:ff}, the kinematic factors in Eq. (\ref{eq:gendoublet}) are given by:
	
	\begin{equation}
		\gamma_\Psi=\gamma_\Phi=\lim_{M_\Phi\rightarrow0}f(M_\Delta,M_\Psi,M_\Phi)=\frac{-i}{(4\pi)^2 M_\Delta M_\Psi}.
	\end{equation}
	
	If the bridge diagram with three heavy propagators is generated, Fig. \ref{fig:ffs}, the contribution to $\alpha_{e\gamma}$ is calculated with:
	\begin{align}
		\gamma_\Psi=&\,f(M_\Delta,M_\Psi,M_\Phi)\,,\\
		\gamma_\Phi=&\,h(M_\Delta,M_\Psi,M_\Phi)\,.
	\end{align}

	\subsection{VLL singlet (triplet) bridge}
	The relevant topologies for the case of a VLL singlet or triplet bridge are the same ones as explored in the previous section for the doublet VLL bridge, changing only the fermionic current since the bridge is now connected to the left-handed muon.
	
	The relevant Lagrangian for an extension of the SM with a triplet, $\Sigma$, that generates the bridge diagram is given by:
	
	\begin{align}
		\label{eq:lag_triplet}
		\mathcal{L}&\supset y_{M} \overline{\ell}_{L}   \sigma^I\phi P_R\Sigma_{I} + y_F  \overline{\Psi}_I \Phi_I e_{R} +T_{KIJ}\left( y_b^R\,  \overline{\Sigma}_K   P_R\Psi_{I} \Phi_J^\dagger+ y_b^L\,  \overline{\Sigma}_K P_L\Psi_{I} \Phi_J^\dagger \right) + \mathrm{h.c.},
	\end{align}
	where $\sigma^I$ denotes  the Pauli matrices and we use the same gauge conventions and general notation introduced in Eqs. (\ref{eq:lagdoublet}) and (\ref{eq:gendoublet}).

	We can write the general result for the bridge contribution to $\alpha_{e\gamma}$ as:
	
	\begin{equation}
		\label{eq:triplet}
		\alpha_{e\gamma}^{2,2}=\frac{-iN e }{4}y_M y_Fy_b^R \sum_{IJ} T_{3IJ} \left[  \gamma_\Psi T^{\gamma,\Psi}_{IJ} +  \gamma_\Phi  T^{\gamma,\Phi}_{IJ} \right]\,.
	\end{equation}

	For the singlet bridge case, $E$, the relevant Lagrangian is the following:
	\begin{align}
		\label{eq:lagsinglet}
		\mathcal{L}&\supset y_M \overline{\ell}_{L} \phi P_R E+ y_F  \overline{\Psi}_I \Phi_I e_{R} +y_b^R\overline{E} P_R\Psi_I \Phi_I^\dagger+ y_b^L \overline{E} P_L\Psi_I \Phi_I^\dagger +\mathrm{ h.c.} \,,
	\end{align}
	where once again we use the same conventions as in Eqs. (\ref{eq:lagdoublet}),(\ref{eq:gendoublet}). The contributions to $\alpha_{e\gamma}$ are given by:
	
	\begin{equation}
		\label{eq:singlet}
		\alpha_{e\gamma}^{2,2}=\frac{iN e}{4}  y_M y_F y_b^R \left(  \tr{[T^{\gamma,\Psi}]}\gamma_\Psi+\tr{[T^{\gamma,\Phi}]}\gamma_\Phi \right)\,. 
	\end{equation}
	
	For the diagram with one heavy fermion and the SM Higgs in the loop, the kinematic factors on both Eqs. (\ref{eq:triplet})  and (\ref{eq:singlet}) and are given by:
	\begin{equation}
		\gamma_\Psi=\gamma_\Phi=\lim_{M_\Phi\rightarrow0}f(M_{E(\Sigma)},M_\Psi,M_\Phi)\,,
	\end{equation}
	whereas for the case in which a heavy fermion and a heavy scalar are in the loop, the kinematic factors read:
	\begin{align}
		\gamma_\psi=&\,f(M_{E(\Sigma)},M_\Psi,M_\Phi)\,,\\
		\gamma_\Phi=&\,h(M_{E(\Sigma)},M_\Psi,M_\Phi)\,.
	\end{align}

	\section{Two-field extensions}
	
	There is a finite number of two-field extensions of the SM which can generate the topologies discussed previously, and as such, we will present the final $\alpha_{e\gamma}$ -- defined in Eq. (\ref{eq:amu}) -- contribution for all of them.
	
	In Table \ref{tab:2fields} we present all the possible completions which in principle generate a contribution to $\alpha_{e\gamma}$ from a bridge topology with only one heavy propagator besides the bridge, Fig. \ref{fig:ff}. Note that once the particle in the bridge is fixed, the quantum numbers of the other heavy particle are also fixed. 
	
	\begin{table}
		\caption{2-field UV completions which generate the bridge topology, with only two heavy propagators. However we consider all possible topologies for the presented $a_\mu$ result both those coming from single field topologies and from the box diagrams.}
		\label{tab:2fields}
		\centering
		\begin{tabular}{llll} 
			\multicolumn{1}{c}{Bridge} &  Other Fermion & $a_\mu$ result  & \\
			\toprule
			\addlinespace[0.25cm]
			\multirow{2}{*}{$E\sim(1,1,-1)$ }& $\Delta\sim(1,2,-1/2)$ & Eq. (\ref{eq:ed}) &   \\
			\addlinespace[0.25cm]
			& $\Delta_3\sim(1,2,-3/2)$ & Eq. (\ref{eq:ed3}) &                                    \\
			\addlinespace[0.5cm]
			\multirow{4}{*}{$\Delta\sim(1,2,-1/2)$ }& $E\sim(1,1,-1)$ & Eq. (\ref{eq:ed}) &                                     \\
			\addlinespace[0.25cm]
			& $\Sigma\sim(1,3,-1)$ & Eq. (\ref{eq:dsig}) &                                     \\
			\addlinespace[0.25cm]
			& $N\sim(1,1,0)$ & Eq. (\ref{eq:dn}) &                                     \\
			\addlinespace[0.25cm]
			& $\Sigma_0\sim(1,3,0)$ & Eq. (\ref{eq:dsig0}) &                                     \\
			\addlinespace[0.5cm]
			
			\multirow{2}{*}{$\Sigma\sim(1,3,-1)$ }& $\Delta\sim(1,2,-1/2)$& Eq. (\ref{eq:dsig}) &                                     \\
			\addlinespace[0.25cm]
			& $\Delta_3\sim(1,2,-3/2)$ & Eq. (\ref{eq:sd3}) &                                    \\
			
			\addlinespace[0.25cm]
			\bottomrule
		\end{tabular}
	\end{table}

	The obtained contributions to $\alpha_{e\gamma}$ are given below, with all results to be understood as divided by the loop factor $(16\pi^2)$:

	\begin{enumerate}[label=\textbf{\arabic*}.]
		
		\item {$E\,\sim(1,1,-1)~\text{and}~\Delta\,\sim(1,2,-1/2)$}
		
		There are two different bridge diagrams that contribute in this case: one with the singlet on the bridge and the doublet in the loop, and vice-versa.
		\begin{equation}
			\label{eq:ed}
			\alpha_{e\gamma}^{2,2}=-\frac{e\, y_M y_F y_b^R}{ 4 M_E M_\Delta }\,.
		\end{equation}
		The couplings above can be interpreted within the Lagrangian in Eq.~\eqref{eq:lagdoublet} or Eq.~\eqref{eq:lagsinglet}.
		
		\item{$E\sim(1,1,-1)$ and $\Delta_3\sim(1,2,-3/2)$}
		
		\begin{equation}
			\label{eq:ed3}
			\alpha_{e\gamma}^{2,2}=-\frac{5 e\, y_M y_F y_b^R}{4 M_E M_{\Delta_3}}\,;
		\end{equation}

		\item{$\Delta\,\sim(1,2,-1/2)$~\text{and}~$\Sigma\,\sim(1,3,-1)$}
		
		There are two bridge diagrams relevant for this case: one with the doublet on the bridge and the triplet in the loop, and vice-versa. 
		\begin{equation}
			\label{eq:dsig}
			\alpha_{e\gamma}^{2,2}=-\frac{9 e\, y_M y_F y_b^R}{4 M_\Delta M_\Sigma }\,.
		\end{equation}
		Both \eqref{eq:lagdoublet} and \eqref{eq:lag_triplet} can be used to interpret this result.
		
		\item{$\Delta\sim(1,2,-1/2)$ and $N\sim(1,1,0)$}
		
		\begin{equation}
			\label{eq:dn}
			\alpha_{e\gamma}^{2,2}=0\,.
		\end{equation}
		This zero has been extensively discussed in the literature in Refs. \cite{Arkani-Hamed:2021xlp,Rose:2022njd}.
		
		\item{$\Delta\sim(1,2,-1/2)$ and $\Sigma_0\sim(1,3,0)$}
		
		\begin{equation}
			\label{eq:dsig0}
			\alpha_{e\gamma}^{2,2}=-\frac{e\, y_M y_F y_b^R}{2 M_\Delta M_{\Sigma_0}}\,;
		\end{equation}

		\item{$\Sigma\sim(1,3,-1)$ and $\Delta_3\sim(1,2,-3/2)$}
		
		\begin{equation}
			\label{eq:sd3}
			\alpha_{e\gamma}^{2,2}=-\frac{5 e\, y_M y_F y_b^R}{4 M_\Sigma M_{\Delta_3} }\,.
		\end{equation}

	\end{enumerate}
	
	These results were cross-checked against those in \cite{Kannike:2011ng, Rose:2022njd, Freitas:2014pua}, and found agreement except for the numerical factors in Eq. (4.8) from Ref.  \cite{Kannike:2011ng} for VLLs with doubly charged components.
	
	A 2-field extension can also generate a bridge topology with 3 heavy propagators, Fig. \ref{fig:ffs}, if the fermion in the bridge is the same as the fermion that propagates in the loop. The extensions for which this can occur are listed in Table \ref{tab:fermionscalar}. The corresponding contributions to $\alpha_{e\gamma}$ are:
	
	\begin{enumerate}[label=\textbf{\arabic*}.]
		
		\item{$E\sim(1,1,-1)$ and $\mathcal{S}_0\sim(1,1,0)$}
		
		\begin{equation}
			\label{eq:es0}
			\alpha_{e\gamma}^{2,2}=-e y_b^{R} y_M y_F\frac{M_E^4-4M_E^2M_{\mathcal{S}_0}^2+3M_{\mathcal{S}_0}^4+2M_{\mathcal{S}_0}^4\llog{[M_E^2/M_{\mathcal{S}_0}^2]}}{4(M_E^2-M_{\mathcal{S}_0}^2)^3}\,;
		\end{equation}
		
		\item{$E\sim(1,1,-1)$ and $\mathcal{S}_2\sim(1,1,-2)$}
		\begin{equation}
			\label{eq:es2}
			\alpha_{e\gamma}^{2,2}=e y_b^{R} y_M y_F\frac{3M_E^4-4M_E^2M_{\mathcal{S}_2}^2+M_{\mathcal{S}_2}^4+(2M_{\mathcal{S}_2}^4-4M_E^2M_{\mathcal{S}_2}^2)\llog{[M_E^2/M_{\mathcal{S}_2}^2]}}{2(M_E^2-M_{\mathcal{S}_2}^2)^3}\,;
		\end{equation}

		\item{$\Delta\sim(1,2,-1/2)$ and $\mathcal{S}_0 \sim (1,1,0)$}
		\begin{equation}
			\label{eq:ds0}
			\alpha_{e\gamma}^{2,2}=-e y_b^{R} y_M y_F \frac{M_\Delta^4-4M_\Delta^2M_{\mathcal{S}_0}^2+3M_{\mathcal{S}_0}^4+2M_{\mathcal{S}_0}^4\llog{[M_\Delta^2/M_{\mathcal{S}_0}^2]}}{4(M_\Delta^2-M_{\mathcal{S}_0}^2)^3} \,;
		\end{equation}
		
		\item{$\Delta\sim(1,2,-1/2)$ and $\mathcal{S}_1 \sim (1,1,-1)$}
		\begin{equation}
			\label{eq:ds1}
			\alpha_{e\gamma}^{2,2}=0\,;
		\end{equation}
		
		\item{$\Delta\sim(1,2,-1/2)$ and $\Xi_0 \sim (1,3,0)$}
		\begin{equation}
			\label{eq:dxi0}
			\alpha_{e\gamma}^{2,2}=e y_b^{R} y_M y_F \frac{M_\Delta^4+4M_\Delta^2M_{\Xi_0}^2-5M_{\Xi_0}^4-(4M_{\Xi_0}^2M_\Delta^2+2M_{\Xi_0}^4)\llog{[M_\Delta^2/M_{\Xi_0}^2]}}{4(M_\Delta^2-M_{\Xi_0}^2)^3} \,;
		\end{equation}
		
		\item{$\Delta\sim(1,2,-1/2)$ and $\Xi_1 \sim (1,3,-1)$}
		\begin{equation}
			\label{eq:dxi1}
			\alpha_{e\gamma}^{2,2}=-e y_b^{R} y_M y_F \frac{7M_\Delta^4-8M_\Delta^2M_{\Xi_1}^2+M_{\Xi_1}^4+(-10M_{\Xi_1}^2M_\Delta^2+4M_{\Xi_1}^4)\llog{[M_\Delta^2/M_{\Xi_1}^2]}}{2(M_\Delta^2-M_{\Xi_1}^2)^3} \,;
		\end{equation}
		
		\item{$\Sigma\sim(1,3,-1)$ and $\Xi_0 \sim (1,3,0)$}
		\begin{equation}
			\label{eq:sigmaxi0}
			\alpha_{e\gamma}^{2,2}=-e y_b^{R} y_M y_F \frac{M_\Sigma^2-M_{\Xi_0}^2+M_{\Xi_0}^2\llog{[M_{\Xi_0}^2/M_\Sigma^2]}}{(M_\Sigma^2-M_{\Xi_0}^2)^2}\,;
		\end{equation}
		
		\item{$\Sigma\sim(1,3,-1)$ and $\Xi_2 \sim (1,3,-2)$}
		\begin{equation}
			\label{eq:sigmaxi2}
			\alpha_{e\gamma}^{2,2}=0\,.
		\end{equation}
		
	\end{enumerate}
	
	A few comments are in order concerning these results. First, we are considering here only one family of heavy particles for simplicity. Therefore, in this type of completions, where the bridge coupling $y_b$ involves two fields that are equal, it will be zero when the gauge structure is antisymmetric. This is the case for completions \eqref{eq:ds1} and \eqref{eq:sigmaxi2}. Note that this is not true in general when dealing with multiple families for heavy fields. On the other hand, for models \eqref{eq:es0}, \eqref{eq:ds0}, \eqref{eq:dxi0} and \eqref{eq:sigmaxi0}, the couplings $y^R_b$ and $y^L_b$ are related by hermitian conjugation; therefore, we redefine $y_b^R\equiv y_b^R+y_b^{L\,*}$ as the effective coupling with the right-handed chirality and write our results using such convention.

	These models have been considered previously in the literature \cite{Kowalska:2017iqv,Calibbi:2018rzv,Freitas:2014pua}, apart from the ones that involve fermion number violating (FNV) vertices which, to the best of our knowledge, are first explored here. However, only the Yukawa-proportional contribution was considered, which made it so that most of the models in Tab. \ref{tab:fermionscalar} were excluded as explanations of $\Delta a_\mu$, since the new particles had to be lighter than what was allowed by experiments. Since we are performing our calculations in the unbroken phase of the SM, it becomes easier to see the chirally enhanced contribution coming from the bridge diagram, which allows for heavier particles to explain $a_\mu$, and as such, opens up this class of models as possible explanations of the anomaly.
	
	A particularly interesting example is the contribution of the model with $\Delta\sim(1,2,-1/2)$ + $\Xi_0\sim(1,3,0)$ extension where the contribution is taken to be always negative in the literature (and as such excluded as an explanation of the observed anomaly). However, from Eq. (\ref{eq:dxi0}), we see that, given the dependence on the couplings, we have the freedom to make the contribution positive in order to account for the observed $\Delta a_\mu$.

	\begin{table}
		\caption{2 field fermion-scalar UV extensions which generate the bridge topology with 3 heavy propagators. Completions in gray color involve fermion number violating interactions. }
		\label{tab:fermionscalar}
		\centering
		\begin{tabular}{lll} 
			Fermion &  Scalar & Result \\
			\toprule
			\addlinespace[0.25cm]
			\multirow{2}{*}{$E\sim(1,1,-1)$ }& $\mathcal{S}_0\sim(1,1,0)$ & Eq. (\ref{eq:es0})   \\
			\addlinespace[0.25cm]
			& $\color{gray} \mathcal{S}_2\sim(1,1,-2)$ & Eq. (\ref{eq:es2})          \\
			\addlinespace[0.5cm]
			\multirow[c]{4}{*}{$\Delta\sim(1,2,-1/2)$ }& $\mathcal{S}_0\sim(1,1,0)$ & Eq. (\ref{eq:ds0})                                \\
			\addlinespace[0.25cm]
			& $\color{gray}\mathcal{S}_1\sim(1,1,-1)$ & Eq. (\ref{eq:ds1})                            \\
			\addlinespace[0.25cm]
			& $\Xi_0\sim(1,3,0)$ & Eq. (\ref{eq:dxi0})                           \\
			\addlinespace[0.25cm]
			& $\color{gray}\Xi_1\sim(1,3,-1)$ & Eq. (\ref{eq:dxi1})                               \\
			\addlinespace[0.5cm]
			\multirow{2}{*}{$\Sigma\sim(1,3,-1)$ } & $\Xi_0\sim(1,3,0)$ & Eq. (\ref{eq:sigmaxi0})                            \\
			\addlinespace[0.25cm]
			& $\color{gray}\Xi_2\sim(1,3,-2)$ & Eq. 
			(\ref{eq:sigmaxi2})                                                              \\
			\addlinespace[0.25cm]
			\bottomrule
		\end{tabular}
	\end{table}
	
	\section{Three-field extensions}
	
	The possible three-field extensions which can generate the bridge diagram of Fig. \ref{fig:ffs} are infinite due to the existence of a heavy loop. As such, in this section we present the conditions set on the gauge representations of the new fields, for the bridge contribution to $a_\mu$ 
	to be generated. These conditions are once again fixed by the particle which runs in the bridge.
	
	All conditions are defined considering no FNV vertices. In the presence of those interactions, one can simply check whether the conjugate version of the fields respects the following conditions.
	
	The extra heavy scalar, $\Phi$, and heavy fermion, $\Psi$, must respect:
	\begin{enumerate}[label=\textbf{\arabic*}.]
		
		\item{VLL singlet}
		\begin{align}
			Y_{\Psi} - Y_\Phi &= -1 \,,\\
			SU(2)_\Phi \otimes SU(2)_{\Psi} &= 1\,;
		\end{align}
		
		\item{VLL doublet}
		\begin{align}
			Y_{\Psi} - Y_\Phi &= -1/2 \,,\\
			SU(2)_\Phi \otimes SU(2)_{\Psi} & = 2\,;
		\end{align}
		
		\item{VLL triplet}
		\begin{align}
			Y_{\Psi} - Y_\Phi &= -1 \,,\\
			\label{eq:su2trip}
			SU(2)_\Phi  \otimes SU(2)_{\Psi}   &= 3\,,\\
			\label{eq:su2sing}
			SU(2)_\Phi  \otimes SU(2)_{\Psi}  &= 1\,,
		\end{align}
		where Eq. (\ref{eq:su2trip}) refers to the coupling with the bridge triplet and Eq. (\ref{eq:su2sing}) to the coupling with the SM right-handed muon.
		
	\end{enumerate}
	
	As for the color charge, we always need to form a singlet with the two fields in the loop, $\Phi$ and $\Psi$. Larger color representations will result in an enhancement factor to the diagram, as explored in \cite{Allwicher:2021jkr} for completions with box diagrams.
	
	Limiting ourselves to, at most, triplet representations of $SU(2)$, the UV completions which can generate the bridge topology are listed in Table \ref{tab:3fields}. 
	\begin{table}
		\caption{Three-field UV extensions which generate the bridge topology, considering only singlet, doublet and triplet $SU(2)$ representations. We show only the $SU(2)$ representations of the 2 extra fields, $\Phi$ and $\Psi$, since the color representation must be the conjugate of each other and their hypercharge must respect the conditions specified above. Switching the assigned $SU(2)$ representations between $\Phi$ and $\Psi$ is also a possible extension.}
		\label{tab:3fields}
		\centering
		\begin{tabular}{ccll} 
			Bridge &  ($SU(2)_\Psi\,$,$\,SU(2)_\Phi$) & Result\\
			\toprule
			\addlinespace[0.25cm]
			\multirow[c]{3}{*}{$E\sim(1,1,-1)$ }& (1,1) & Eq. \eqref{eq:E11} \\
			\addlinespace[0.25cm]
			& (2,2) &  Eq. \eqref{eq:E22} \\
			\addlinespace[0.25cm]
			& (3,3) &  Eq. \eqref{eq:E33} \\
			\addlinespace[0.5cm]
			\multirow[c]{2}{*}{$\Delta\sim(1,2,-1/2)$ }& (2,1) & Eqs. \eqref{eq:D21}, \eqref{eq:D12}                            \\
			\addlinespace[0.25cm]
			& (2,3) & Eqs. \eqref{eq:D23}, \eqref{eq:D32}   \\
			\addlinespace[0.5cm]
			\multirow[c]{2}{*}{$\Sigma\sim(1,3,-1)$ }& (2,2) & Eq. \eqref{eq:T22}  \\
			\addlinespace[0.25cm]
			& (3,3) &  Eq. \eqref{eq:T33}\\
			
			\addlinespace[0.25cm]
			\bottomrule
		\end{tabular}
	\end{table}
	
	It is evident that none of these completions is among the ones explored in Refs. \cite{Calibbi:2018rzv,Calibbi:2019bay,Arnan:2019uhr,Crivellin:2021rbq,Allwicher:2021jkr,Arcadi:2021cwg} and as such constitute novel extensions of the SM which can in principle contribute to $a_\mu$. Let us present the final expression for these models, using the notation introduced in Eqs. \eqref{eq:lagdoublet}, \eqref{eq:lag_triplet} and \eqref{eq:lagsinglet}. Again, a factor $1/16\pi^2$ is omitted in all cases and we neglect contributions proportional to the lepton Yukawa. We present the results in terms of a generic hypercharge for $\Psi$, $Y_\Psi$, and use the notation $(\Psi,\Phi)\sim(SU(2)_\Psi,SU(2)_\Phi)$ to indicate the $SU(2)$ representations of the fields, as listed in Table \ref{tab:3fields}.
	
	\begin{enumerate}[label=\textbf{\arabic*}.]
		\item{$E\sim(1,1,-1) + (\Psi,\Phi)\sim(1,1)$}
		\begin{align}
			\label{eq:E11}
			\begin{split}
				\alpha_{e\gamma}^{2,2}=\frac{e N M_\Psi y_{M} y_F y_b^R}{4M_E(M_\Psi^2-M_\Phi^2)^3}\bigg\{
				(M_\Psi^2-M_\Phi^2)(M_\Phi^2(1-2Y_\Psi)+M_\Psi^2(1+2Y_\Psi))\bigg.\\
				\bigg.-2(-M_\Phi^4 Y_\Psi+M_\Psi^2M_\Phi^2(1+Y_\Psi))\llog{[M_\Psi^2/M_\Phi^2]} \bigg\}\,;
			\end{split}
		\end{align}  
		
		\item{$E\sim(1,1,-1) + (\Psi,\Phi)\sim(2,2)$}
		\begin{equation}
			\label{eq:E22}
			\begin{split}
				\alpha_{e\gamma}^{2,2}=
				\frac{e N M_\Psi y_{M} y_F y_b^R}{2M_E(M_\Psi^2-M_\Phi^2)^3}
				\bigg\{(M_\Psi^2-M_\Phi^2)(M_\Phi^2(1-2Y_\Psi)+M_\Psi^2(1+2Y_\Psi))\\
				-2(-M_\Phi^4 Y_\Psi+M_\Psi^2M_\Phi^2(1+Y_\Psi))\llog{[M_\Psi^2/M_\Phi^2]}\bigg\}\,;
			\end{split}
		\end{equation}  
		
		\item{$E\sim(1,1,-1) + (\Psi,\Phi)\sim(3,3)$}
		\begin{equation}
			\label{eq:E33}
			\begin{split}
				\alpha_{e\gamma}^{2,2}=
				\frac{3e N M_\Psi y_{M} y_F y_b^R}{4M_E(M_\Psi^2-M_\Phi^2)^3}
				\bigg\{(M_\Psi^2-M_\Phi^2)(M_\Phi^2(1-2Y_\Psi)+M_\Psi^2(1+2Y_\Psi))\\-2(-M_\Phi^4 Y_\Psi+M_\Psi^2M_\Phi^2(1+Y_\Psi))\llog{[M_\Psi^2/M_\Phi^2]}\bigg\}\,;
			\end{split}
		\end{equation}

		\item{$\Delta\sim(1,2,-1/2) + (\Psi,\Phi)\sim(2,1)$}
		\begin{equation}
			\label{eq:D21}
			\begin{split}
				\alpha_{e\gamma}^{2,2}=
				\frac{e N M_\Psi y_{M} y_F y_b^R}{4M_\Delta(M_\Psi^2-M_\Phi^2)^3}
				\bigg\{2(M_\Psi^2-M_\Phi^2)(M_\Phi^2(1-Y_\Psi)+M_\Psi^2 Y_\Psi)\\-(M_\Phi^4(1-2Y_\Psi)+M_\Psi^2 M_\Phi^2(1+2Y_\Psi))\llog{[M_\Psi^2/M_\Phi^2]}\bigg\}\,;
			\end{split}
		\end{equation}  
		
		\item{$\Delta\sim(1,2,-1/2) + (\Psi,\Phi)\sim(1,2)$}
		\begin{equation}
			\label{eq:D12}
			\begin{split}
				\alpha_{e\gamma}^{2,2}=
				\frac{e N M_\Psi y_{M} y_F y_b^R}{4M_\Delta(M_\Psi^2-M_\Phi^2)^3}
				\bigg\{(M_\Psi^2-M_\Phi^2)(M_\Phi^2(1-2Y_\Psi)+M_\Psi^2 (1+2Y_\Psi))\\-2(-M_\Phi^4Y_\Psi+M_\Psi^2 M_\Phi^2(1+Y_\Psi))\llog{[M_\Psi^2/M_\Phi^2]}\bigg\}\,;
			\end{split}
		\end{equation}  
		
		\item{$\Delta\sim(1,2,-1/2) + (\Psi,\Phi)\sim(2,3)$}
		\begin{equation}
			\label{eq:D23}
			\begin{split}
				\alpha_{e\gamma}^{2,2}=
				\frac{e N M_\Psi y_{M} y_F y_b^R}{4M_\Delta(M_\Psi^2-M_\Phi^2)^3}
				\bigg\{2(M_\Psi^2-M_\Phi^2)(M_\Phi^2(1-3Y_\Psi)+M_\Psi^2 (2+3Y_\Psi))\\+(M_\Phi^4(1+6Y_\Psi)-M_\Psi^2 M_\Phi^2(7+6Y_\Psi))\llog{[M_\Psi^2/M_\Phi^2]}\bigg\}\,;
			\end{split}
		\end{equation}  
		
		\item{$\Delta\sim(1,2,-1/2) + (\Psi,\Phi)\sim(3,2)$}
		\begin{equation}
			\label{eq:D32}
			\begin{split}
				\alpha_{e\gamma}^{2,2}=
				-\frac{e N M_\Psi y_{M} y_F y_b^R}{4M_\Delta(M_\Psi^2-M_\Phi^2)^3}
				\bigg\{(M_\Psi^2-M_\Phi^2)(M_\Phi^2(7-6Y_\Psi)+M_\Psi^2 (-1+6Y_\Psi))\\-2(M_\Phi^4(2-3Y_\Psi)+M_\Psi^2 M_\Phi^2(1+3Y_\Psi))\llog{[M_\Psi^2/M_\Phi^2]} \bigg\}\,;
			\end{split}
		\end{equation}

		\item{$\Sigma\sim(1,3,-1) + (\Psi,\Phi)\sim(2,2)$}
		\begin{equation}
			\label{eq:T22}
			\begin{split}
				\alpha_{e\gamma}^{2,2}=
				-\frac{e N M_\Psi y_{M} y_F y_b^R}{2M_\Sigma(M_\Psi^2-M_\Phi^2)^2}
				\bigg\{M_\Psi^2-M_\Phi^2-M_\Phi^2\llog{[M_\Psi^2/M_\Phi^2]}\bigg\}\,;
			\end{split}
		\end{equation} 
		
		\item{$\Sigma\sim(1,3,-1) + (\Psi,\Phi)\sim(3,3)$}
		\begin{equation}
			\label{eq:T33}
			\begin{split}
				\alpha_{e\gamma}^{2,2}=
				-\frac{e N M_\Psi y_{M} y_F y_b^R}{M_\Sigma(M_\Psi^2-M_\Phi^2)^2}
				\bigg\{M_\Psi^2-M_\Phi^2-M_\Phi^2\llog{[M_\Psi^2/M_\Phi^2]}\bigg\}\,.
			\end{split}
		\end{equation} 
		
	\end{enumerate}
	
	\section{General phenomenological considerations}
	
	Among the new particles introduced, the one expected to be the most constrained by experiment is the VLL in the bridge, as it is the only one which must have a coupling to 2 SM particles.
	
	The mixing of VLLs with the SM muon is bounded by electroweak precision observables (EWPO) \cite{deBlas:2013gla,Crivellin:2020ebi}:
	\begin{align}
		\label{eq:mixing}
		\frac{v}{M_E}y_M &\lesssim 0.03\, (0.04)\,, \\
		\frac{v}{M_\Delta}y_M &\lesssim 0.065\,(0.075)\,, \\
		\frac{v}{M_\Sigma}y_M &\lesssim 0.1\,(0.11)\,,
	\end{align}
	for the singlet, doublet and triplet of $SU(2)$ respectively at 1 (2) $\sigma$ confidence levels.
	
	Direct searches at colliders for these VLLs set lower limits on their masses. Ref. \cite{ATLAS:2015qoy} set a limit of $M_E\gtrsim175\,\mathrm{GeV}$ for a singlet VLL decaying through muon or electron channels. However, Ref. \cite{Guedes:2021oqx} estimates that the HL phase of the LHC could exclude masses lighter than $800\;\mathrm{GeV}$. The doublet has more recently been probed by CMS \cite{CMS:2019hsm}, where masses below $\sim800\,\mathrm{GeV}$ are excluded; these are conservative results for our study as only tau decays were considered, whereas in order to explain $a_\mu$ the VLL must couple to muons. For the triplet case, Ref. \cite{Ashanujjaman:2022cso} estimates the discovery reach of the LHC at $3\,\mathrm{ab}^{-1}$ at $5\,\sigma$ to be of approximately $1.4 \;\mathrm{TeV}$.
	
	Other direct bounds could be taken from the particles that run in the loop, through EW or QCD (when possible) pair production. However, this -- and other possible probes -- are clearly dependent on the choice of model and it is beyond the scope of this paper to go over the particular phenomenological implications of every model. 
	
	Another common issue with chirally enhanced solutions to $a_\mu$ has to do with the large contribution to the muon Yukawa through the same diagrams which explain $a_\mu$ but without the gauge boson insertion. This is not a problem for the bridge diagram with a VLL triplet since, in order to have a non-zero contribution, an insertion of a $W$-boson is needed. For the case of the singlet and doublet VLL bridges, indeed one can expect a sizable contribution to the muon Yukawa; however, this could only be bounded by fine-tuning arguments to account for a cancellation between this contribution and a possible tree-level coefficient. One can further explore possible UV scenarios in which the Yukawa couplings and the dipole operators have the same origin \cite{Baker:2021yli,Greljo:2021npi,Yin:2021yqy}.
	
	\section{The triple triplet model}
	
	So far, we have focused only on the contributions of SM extensions towards $a_\mu$. However, given the degrees of freedom we introduce by considering multi-field extensions, we have enough flexibility to try to accommodate other anomalous observations. The purpose of this section is to show a particular example of how the bridge topology can connect explanations of different anomalies.
	
	Of particular interest within the latest measurements is the suggestion of lepton flavour universality violation in $B$-meson decays. The ratio
	\begin{equation}
		R_K^{(*)} = \frac{\mathrm{BR}(B\rightarrow K^{(*)} \overline{\mu} \mu)}{\mathrm{BR}(B\rightarrow K^{(*)} \overline{e} e)}
	\end{equation}
	is close to 1 in the SM, but the observations of these ratios show a combined deviation from the SM prediction of more than $4\,\sigma$ \cite{LHCb:2017avl,LHCb:2019hip,Alguero:2019ptt,Aebischer:2019mlg,Ciuchini:2019usw}.
	
	Another tension within the SM is the possible violation of unitarity in the first row of the CKM matrix, known as the Cabibbo Angle Anomaly (CAA). Direct measurements of $V_{us}$ from leptonic kaon decays seem to be in tension with those which assume CKM unitarity coming from super-allowed $\beta$ decays \cite{Grossman:2019bzp,Belfatto_2020}. The significance of this tension is quoted as $3$ or $5\,\sigma$ depending on the parametrization of the $\beta$ decays. We parametrize this tension as the deviation from unity of the $R(V_{us})$ observable, which is defined as the ratio between the extraction of $V_{us}$ from purely muonic kaon decays over the one extracted from $\beta$ decays assuming unitarity \cite{Crivellin:2020lzu,Kirk:2020wdk}. 
	In \cite{Crivellin:2020lzu,Kirk:2020wdk} it is shown how this deviation is directly related to a correction in the muonic vertex with the $W$ boson, which we denote by $\epsilon_{\mu\mu}$:
	\begin{equation}
	  \mathcal{L} \supset \frac{g}{\sqrt{2}}W_\mu^{-} \overline{\ell}_i \gamma^\mu P_L \nu_j (\delta_{ij} + \epsilon_{ij}),
	\end{equation}
	where $i,j$ run over lepton families. 
	
	With the goal of also explaining these two anomalies, we consider one specific realization of the last class of models in Table \ref{tab:3fields}, in which the SM is extended with the vector-like lepton triplet,  $\Sigma$, a triplet scalar leptoquark with hypercharge $-1/3$, $S_3$, and a triplet vector-like quark, $\Psi_Q$, with hypercharge $-4/3$. 
	
	The triplet leptoquark is well-known to be a good solution for the neutral flavour anomalies at tree-level \cite{Gherardi:2020qhc,Angelescu:2021lln,Kumar:2018kmr,Crivellin_2021}, whereas the vector-like lepton triplet can explain the CAA, in spite of creating some tension with EWPO \cite{Crivellin:2020ebi,Kirk:2020wdk}\footnote{The latest CDF II measurement of the $W$-boson mass \cite{doi:10.1126/science.abk1781} has significantly increased the tension between CAA and EWPO (see \cite{Blennow:2022yfm} for a recent analysis). However, addressing this anomaly is beyond the scope of this paper.}. In order to explain $a_\mu$ with the bridge topology including these two particles, we fix the quantum numbers of $\Psi_Q$.
	
	The Lagrangian of this model is the following:
	\begin{align}
		\begin{split}
			\label{lag_model_S3}
			\mathcal{L} &\supset y_{T}^{i} \overline{\ell}_{Li} \phi  \sigma^I \Sigma^{I}_{R} + y_Q^{i} \overline{\Psi}^I_{QL} S_3^I \,e_{Ri} + i y_b^L \epsilon^{IJK} \overline{\Sigma}^I_R \Psi_{Q,L}^J S_3^{K\dagger} + i y_b^R \epsilon^{IJK} \overline{\Sigma}^I_L \Psi_{Q,R}^J S_3^{K\dagger} \\
			&+ \lambda_S^{ij} \overline{q}^c_{Li} i \sigma^2 \sigma^I \ell_{Lj} S_3^{I\dagger} + \lambda_U^i \overline{u_R}_i\,\Sigma^{c, I} S_3^I + \mathrm{h.c.},
		\end{split}
	\end{align}
	with $q^c\equiv \mathcal{C}\overline{q}^T$ and $\mathcal{C}$ the charge conjugation matrix. 
	The SM $SU(2)$ doublets $q_L$ and $\ell_L$ are in the down-quark and charged lepton diagonal basis, respectively.
	
	In the simplest version of this model, we will consider the minimal set of couplings which allow for an explanation of the observed discrepancies in $B$-meson decays, $V_{\mathrm{CKM}}$ unitarity and the anomalous magnetic moment of the muon. Therefore we suppose just one family of heavy particles and assume that new physics only couples to second generation leptons. This also helps to avoid some constraints like the Lepton Flavour Violating (LFV) decay $\mu\rightarrow e\gamma$. With respect to quarks, we will only allow for second and third generation couplings in $\lambda_S^{ij}$, namely $\lambda_S^{s\mu}$ and $\lambda_S^{b\mu}$.
	
	The aforementioned anomalies are explained in this model by the generation of $\mathcal{O}_{\ell q}^{(1),(3)}$ at tree-level by $S_3$ exchange, $\mathcal{O}_{H\ell}^{(3)}$ also at tree-level by $\Sigma$ exchange and a bridge-like one-loop contribution to $\Delta a_\mu$. The expressions for the relevant Wilson coefficients are: 
	\begin{equation}
		[\alpha_{\ell q}^{(1)}]_{i,j,k,l}=\frac{3\lambda_S^{*ki}\lambda_S^{lj}}{4M_{S_3}^2}+\mathcal{O}(\frac{1}{16\pi^2})\,,
	\end{equation}
	\begin{equation}
		[\alpha_{\ell q}^{(3)}]_{i,j,k,l}=\frac{\lambda_S^{*ki}\lambda_S^{lj}}{4M_{S_3}^2}+\mathcal{O}(\frac{1}{16\pi^2})\,,
	\end{equation}
	\begin{equation}
		[\alpha_{H\ell}^{(3)}]_{i,j}=\frac{y_T^i y_T^{*j}}{4 M_\Sigma^2}+\mathcal{O}(\frac{1}{16\pi^2})\,,
	\end{equation}
	\begin{equation}
		[\alpha_{eB}]_{i,j}\simeq 0\,,
	\end{equation}
	\begin{equation}
		[\alpha_{eW}]_{i,j}\simeq \frac{3 g_W y_b^R \, y_T^i y_Q^j}{16\pi^2} \frac{M_{\Psi_Q}}{M_\Sigma}\left(\frac{M_{\Psi_Q}^2-M_{S_3}^2+M_{S_3}^2 \,\llog{\left[\frac{M_{S_3}^2}{M_{\Psi_Q}^2}\right]}}{(M_{\Psi_Q}^2-M_{S_3}^2)^2}\right)\,,
	\end{equation}
	where the $\simeq$ means that we are neglecting Yukawa-suppressed contributions and the notation of the Wilson coefficients follows the convention of {\tt matchmakereft} \cite{Carmona:2021xtq}.
	
	Explaining $R^{(*)}_K$ and CAA essentially fixes the ratios $x_S\equiv\lambda_S^{*s\mu}\lambda_S^{b\mu}/M_{S_3}^2$ and $x_T\equiv y_T^\mu/M_\Sigma$, up to one-loop corrections that break this scale invariance in couplings over masses. However, the loop factor suppression assures that observables are approximately flat on the values of the masses (in a certain range). The $x_T$ ratio also enters in the expression for $\Delta a_\mu$, but we have enough freedom with the couplings $y_b^R$ and $x_F\equiv y_Q^\mu/M_{\Psi_Q}$ to fix both observables to the desired value. Note also that, since they couple two and three heavy fields, both $y_b$ and $x_F$ always generate coefficients at one-loop order, so in principle one expects a wider parameter space in comparison to other couplings.
	
	In order to study the one-loop low-energy phenomenology of the model, we computed the complete one-loop matching with {\tt matchmakereft} \cite{Carmona:2021xtq}\footnote{The output from {\tt matchmakereft} is provided in an auxiliary file.} and used {\tt smelli} \cite{Aebischer_2019,Stangl:2020lbh,Aebischer:2018bkb,Straub:2018kue} to construct a $\chi^2$ function from the observables at low energy in terms of the Wilson coefficients matched at a high scale. We then performed a fit, minimizing the $\chi^2$ function using the {\tt iminuit} \cite{iminuit} python package; for this fit we considered the observables given in {\tt smelli} in the classes of leptonic observables (which include magnetic dipole moments for leptons), lepton flavour universality for neutral currents (for anomalies in $B$ decays such as $R^{(*)}_K$), EWPO (which contain observables sensitive to deviations in the electroweak vertices) and quark flavour related observables (which include meson decays and mixing)\footnote{Further details on all the observables included in these classes can be found in Appendix D of \cite{Aebischer_2019} where they are listed.}.  Besides these observables taken directly from {\tt smelli}, we also added $\epsilon_{\mu\mu}$.
	
	In addition to the flavour assumptions commented above,
	we further imposed the couplings to be lower than 1, and
	fixed the masses to $M_\Sigma=3.4$ TeV, $M_{S_3}=2$ TeV and $M_{\Psi_Q}=4.6$ TeV. As explained, the observables were essentially flat in masses between 1-5 TeV, so we chose this hierarchy as an example that avoided current experimental detection limits but that could be reached by upcoming searches. Other hierarchies and values for the masses between 1-5 TeV are also feasible and yield similar results.
	
	The best fit point in this setup is:
	\begin{equation}
		\begin{aligned}[c]
			&x_F=0.2\;\tev{}^{-1},\\
			&x_T=0.17\;\tev{}^{-1},\\
			&y_b^L=0.10,
		\end{aligned}
		\qquad\qquad
		\begin{aligned}[c]
			&x_S=0.00078\;\tev{}^{-2},\\
			&\lambda_S^{b\mu}=0.07,\\
			&y_b^R=0.13,
		\end{aligned}
	\end{equation}
	which corresponds to a global pull from the SM of 6.2 $\sigma$. For the calculation of this pull, we considered the observables that were fitted, i.e., the ones available in {\tt smelli} in the classes EWPO, leptonic observables, lepton flavour universality for neutral currents and quark flavour observables; we do not include $\epsilon_{\mu\mu}$ since we did not consider its correlations with the observables in the stated classes. In Table \ref{tab:pulls}, we collect some of the individual pulls, both from experiment and SM, for the most relevant observables.

	\begin{table}
		\caption{Individual values for the SM prediction, model prediction, experimental measure and pulls of the most relevant observables as given by {\tt smelli}. Definitions of these observables (and updated values) can be read from the {\tt flavio} \cite{Straub:2018kue} documentation, which {\tt smelli} uses to calculate contributions from the Wilson coefficients to low-energy observables.} 
		\label{tab:pulls}
		\centering
	\resizebox{\textwidth}{!}{	\begin{tabular}{cccccccc} 
			Observable & \thead{SM\\Prediction}& \thead{Model\\Prediction}&Experiment& \thead{Pull\\model ($\sigma$)}& Pull SM ($\sigma$)\\
			\toprule
			\addlinespace[0.25cm]
			$a_\mu$&0.0011659181(4)&0.0011659201(4)&0.0011659206(4)&0.82&4.22\\
			$\langle R_{\mu e} \rangle(B^{\pm}\rightarrow K^{\pm}\ell^+\ell^-)^{[1.0,6.0]}$&1&0.79&0.85(5)&1.41&3.21\\
			$\langle R_{\mu e} \rangle(B^{0}\rightarrow K^{*0}\ell^+\ell^-)^{[0.045,1.1]}$&0.93&0.87&0.65(12)&1.98&2.39\\
			$\langle R_{\mu e} \rangle(B^{0}\rightarrow K^{*0}\ell^+\ell^-)^{[1.1,6.0]}$&0.99&0.79&0.68(12)&1.04&2.55\\
			$\epsilon_{\mu\mu}$&0&0.40e-3&0.58(15)e-3&1.20&3.87\\
			$\Delta M_s$&1.25(8)e-11&1.25(8)e-11&1.1688(14)e-11&1.08&1.07\\
			$\Delta M_d$&3.9(5)e-13&3.9(5)e-13&3.33(15)e-13&1.25&1.25\\
			$M_W$&80.36&80.35&80.379(12)&2.28&1.72\\
			$A_e$&0.147&0.146&0.151(2)&2.77&2.22\\

			\bottomrule
		\end{tabular}}
	\end{table}

\begin{figure}
		\centering
		\includegraphics[width=0.75\textwidth]{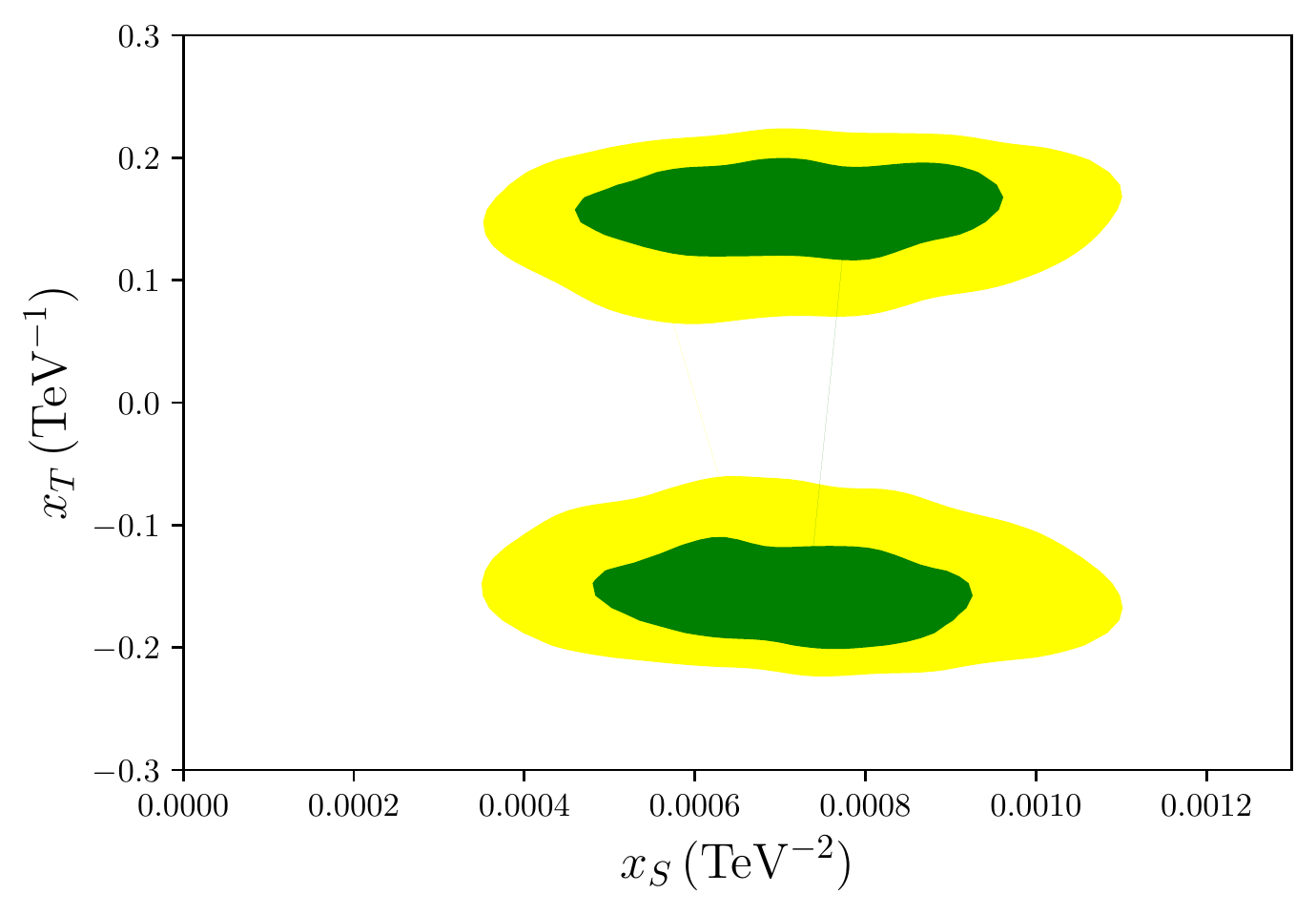}
		\caption{The 1 (2)-$\sigma$ regions in green (yellow) around the model's best fit point. For each $x_S$ and $x_T$ point in the plot, the other couplings were marginalized in order to minimize the $\chi^2$. The observables included in the fit were the ones available in {\tt{smelli}} in the classes EWPO, leptonic observables, lepton flavour universality for neutral currents and quark flavour observables, and $\epsilon_{\mu\mu}$.}
		\label{fig:fit}
	\end{figure}

\begin{figure}
		\centering
		\includegraphics[width=0.65\textwidth]{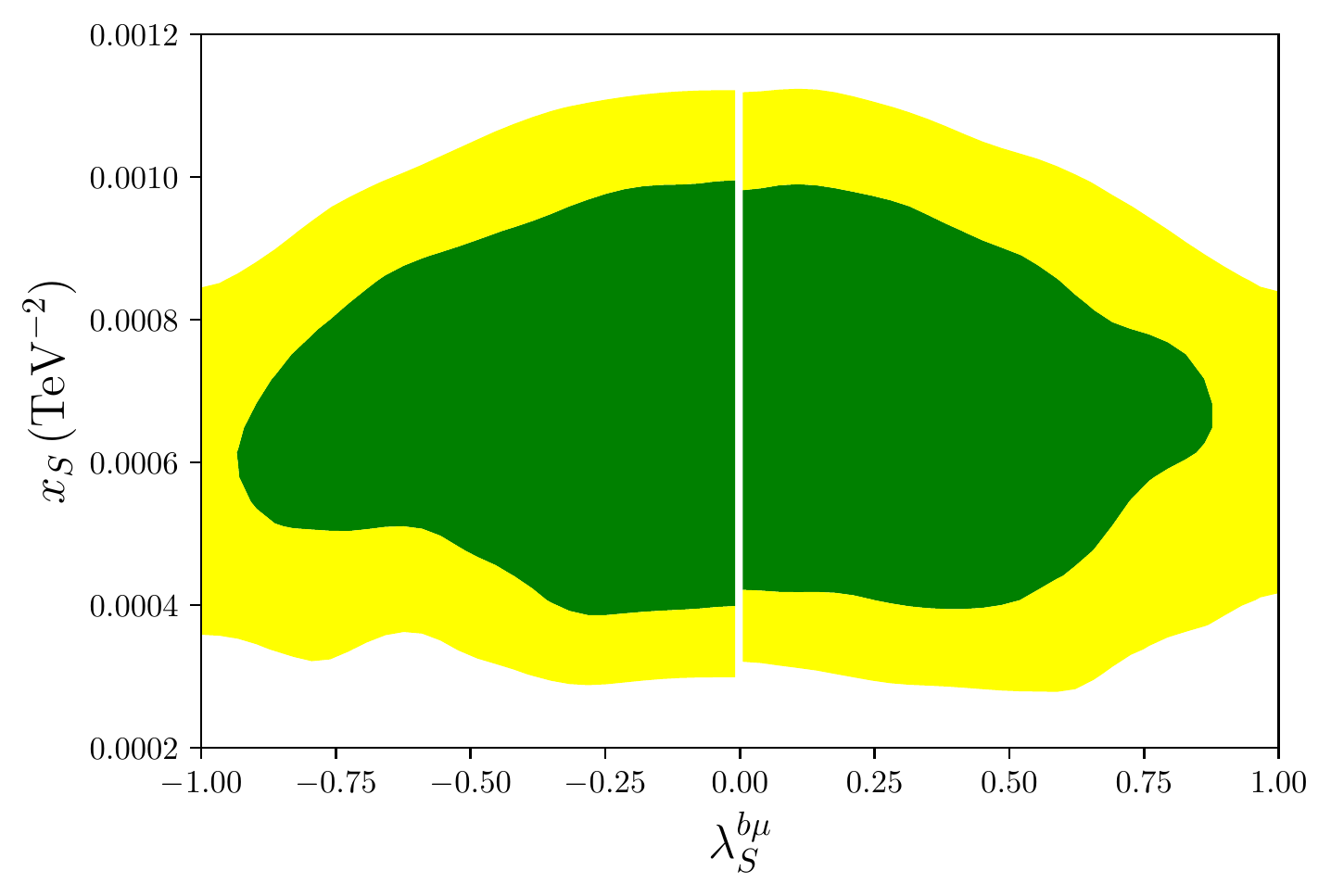}
		\caption{The 1 (2)-$\sigma$ regions in green (yellow) around the model's best fit point. For each point in the plot, the other couplings were marginalized in order to minimize the $\chi^2$. The observables included in the fit were the ones available in {\tt{smelli}} in the classes EWPO, leptonic observables, lepton flavour universality for neutral currents and quark flavour observables, and $\epsilon_{\mu\mu}$. Values of $\lambda_S^{b\mu}$ very close to zero were not plotted because that would imply $\lambda_S^{s\mu}$ larger than 1 for a fixed $x_S$ and $M_{S_3}$.}
		\label{fig:fit_lamb}
	\end{figure}

\begin{figure}
		\centering
		\begin{subfigure}[b]{0.49\textwidth}
         \centering
         \includegraphics[width=\textwidth]{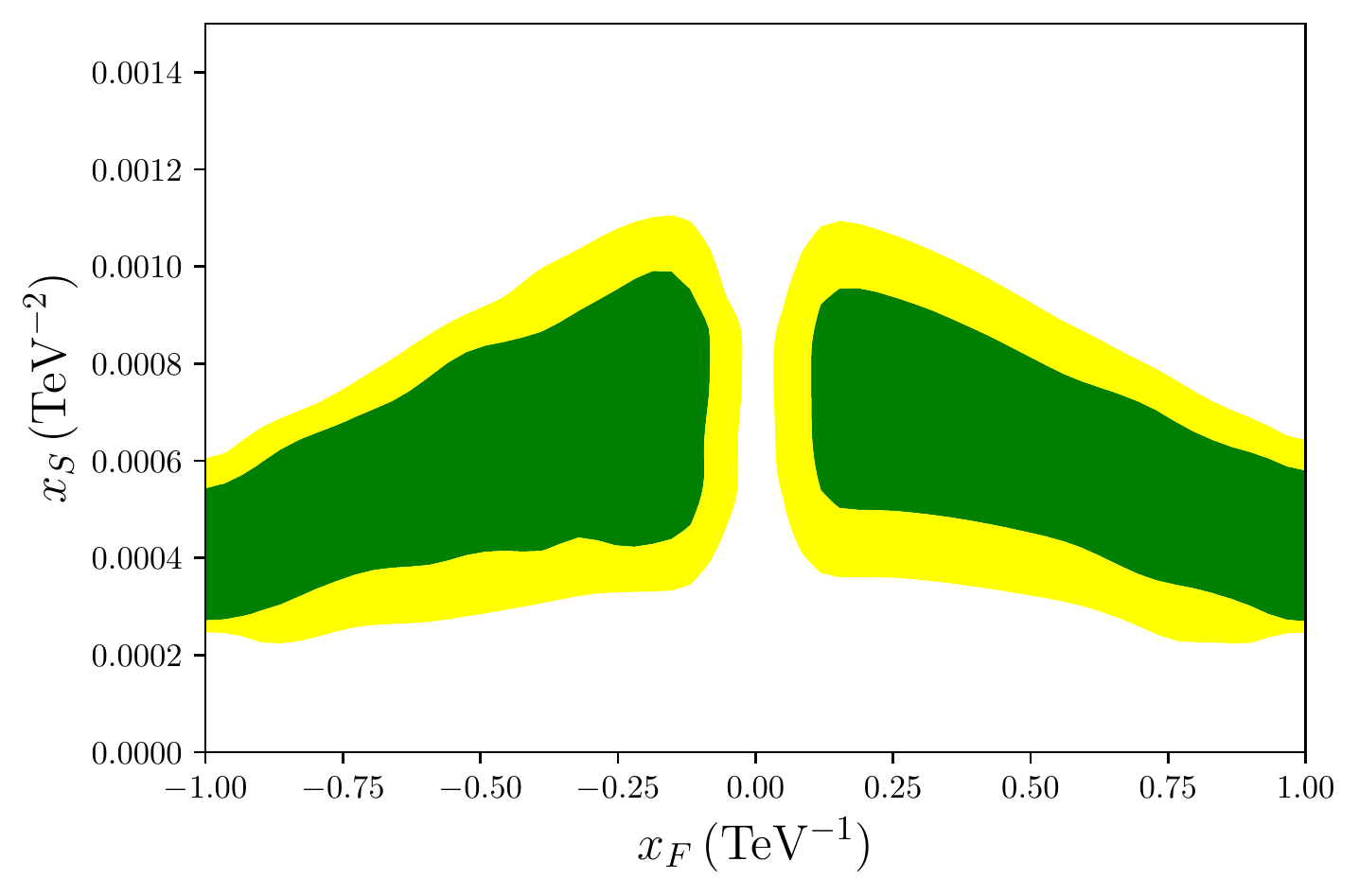}

        \end{subfigure}
        \hfill
        \begin{subfigure}[b]{0.49\textwidth}
         \centering
         \includegraphics[width=\textwidth]{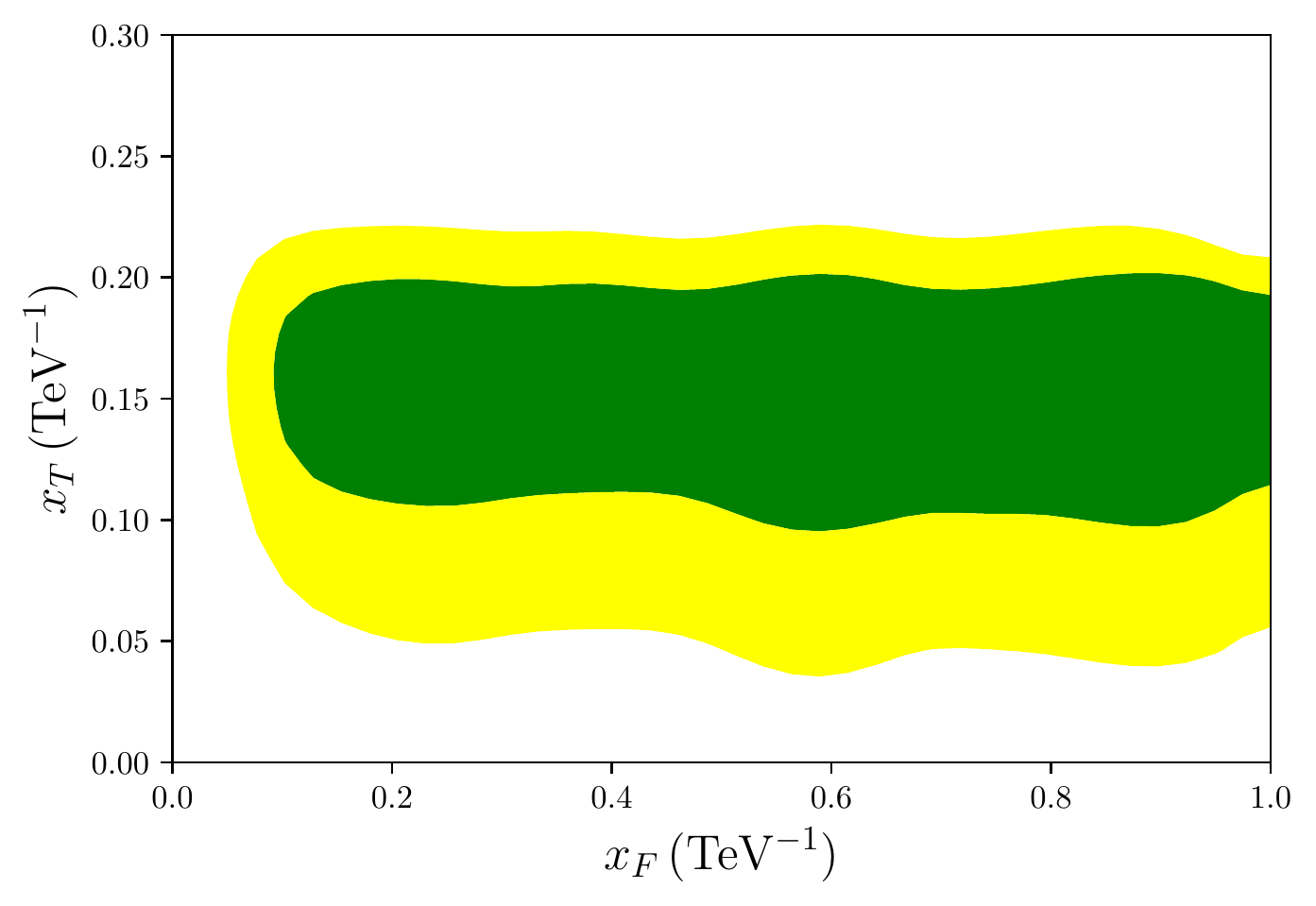}

        \end{subfigure}
        
        \begin{subfigure}[b]{0.49\textwidth}
         \centering
         \includegraphics[width=\textwidth]{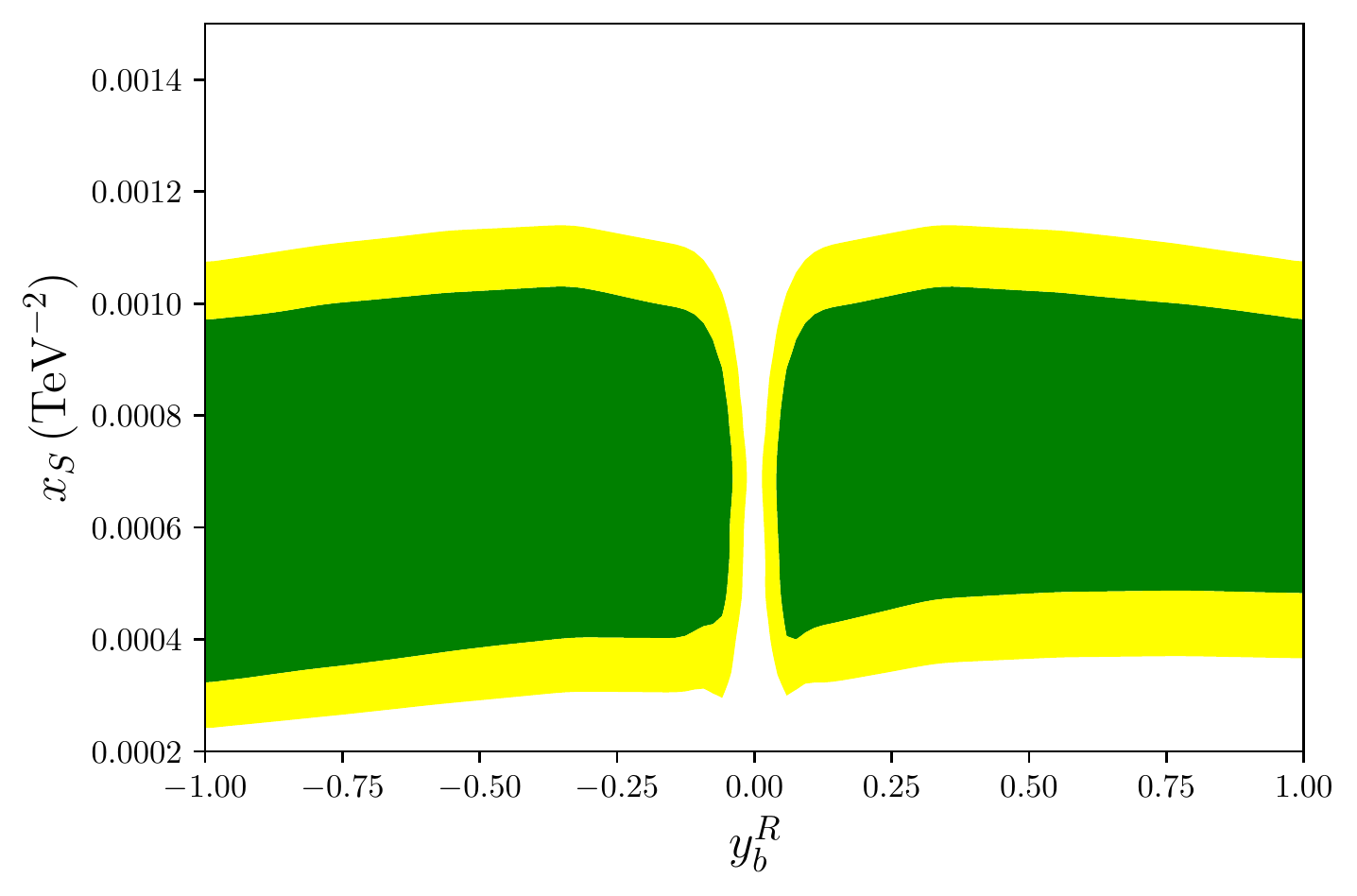}
       
        \end{subfigure}
        \hfill
        \begin{subfigure}[b]{0.49\textwidth}
         \centering
         \includegraphics[width=\textwidth]{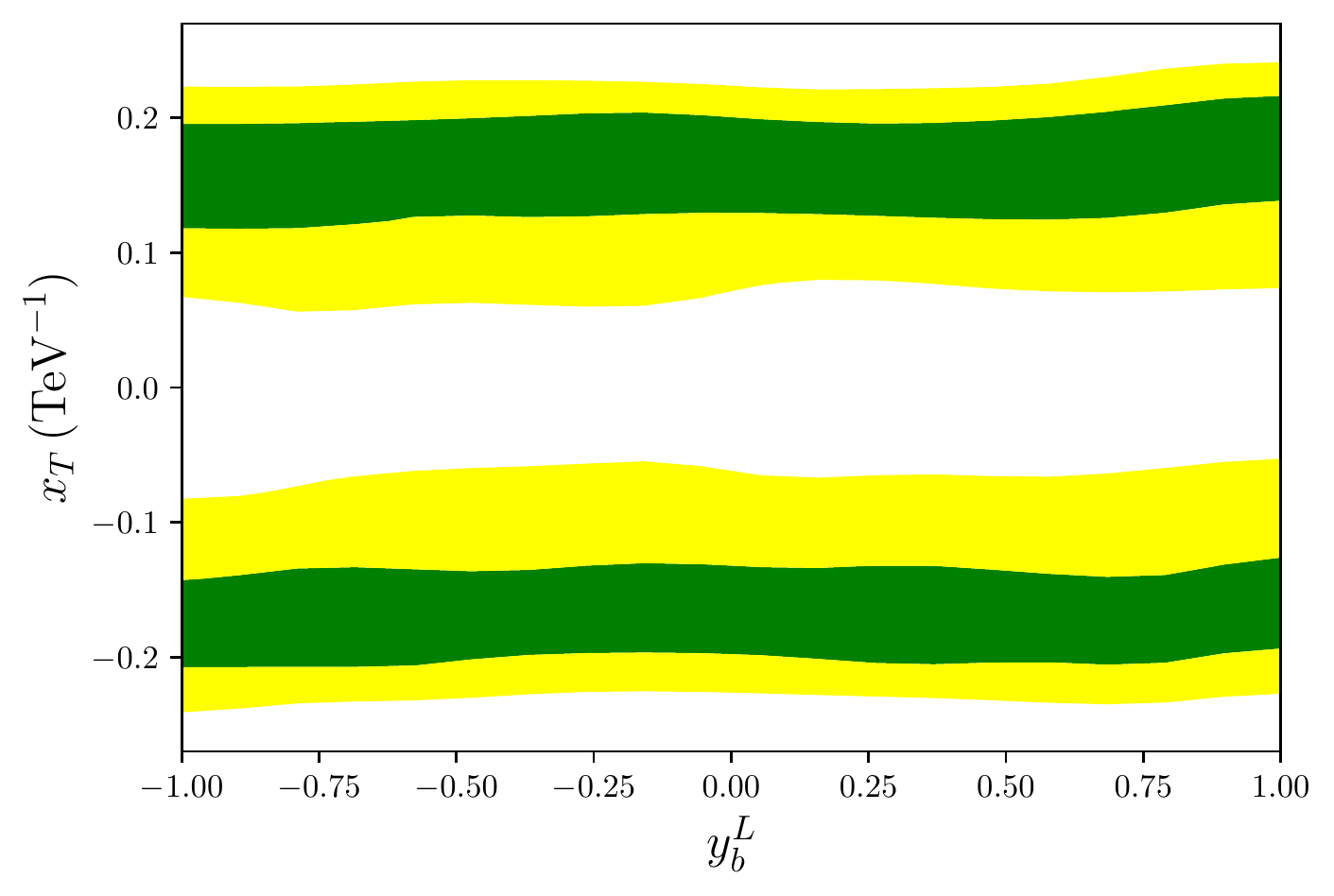}
        \end{subfigure}
	    \caption{The 1 (2)-$\sigma$ regions in green (yellow) around the model's best fit point. For every plot, the vertical axis represents a coupling that enters observables a tree-level, whereas the horizontal axis represents one that only contributes at one-loop. For each point in the plot, the other couplings were marginalized in order to minimize the $\chi^2$. The observables included in the fit were the ones available in {\tt{smelli}} in the classes EWPO, leptonic observables, lepton flavour universality for neutral currents and quark flavour observables, and $\epsilon_{\mu\mu}$.}
	    \label{fig:fit_loop}
	\end{figure}

	In Figs. \ref{fig:fit} and \ref{fig:fit_lamb} we show the 1- and 2-$\sigma$ regions from the best-fit point for the model parameters that generate tree-level Wilson coefficients using the global likelihood provided by {\tt smelli}. For each point in the grid, we minimize the value of $\chi^2$ by varying the other parameters of the model. The profiles in the rest of the variables that contribute only at one-loop are very similar to what one would expect from only taking the tree level solutions (i.e. they feature somewhat flat directions), showing that the one-loop effects give us enough freedom to explain $\Delta a_\mu$ without spoiling the anomalies independently explained at tree level. These flat directions are illustrated in the plots of Fig. \ref{fig:fit_loop}, where the coupling on the $x$-axis only contributes at one-loop; note that a vanishing $y_b^L$ coupling is allowed since it does not contribute to the relevant anomalies.

	\section{Conclusions}
	We have studied in detail the chirally enhanced contribution to $a_\mu$ arising from the bridge topology shown in Fig. \ref{fig:ffs}. Results for this contribution have been presented for arbitrary extensions of the SM, in terms of the gauge representations of the heavy fields. 
	
	The particular representations of the combination of heavy fields which can generate the bridge diagram have been classified in terms of 2- or 3-field extensions. The full results for the contribution to $a_\mu$ of all possible 2-field extensions have been presented; for 3-field extensions, due to the infinite number of possibilities, we presented the results for extensions with singlet, doublet and triplet representations $SU(2)$. For higher representations, the general results presented in Section 3 (and in Appendix A for the box diagrams) can be used to compute the contribution to $a_\mu$.
	
	Within these results, we arrive at a class of 2-field extensions with a fermion and a scalar which had previously been discarded in the literature but that, with this chirally enhanced contribution from the bridge diagram, can in principle be viable solutions to the $a_\mu$ anomaly. The studied 3-field extensions which generate the bridge topology also represent a completely new class of models to explain the $a_\mu$ anomaly.
	
	An interesting avenue to pursue is to study the specific phenomenology of these models to explore the parameter space in which they can explain $a_\mu$ and also other anomalous observations. For example, within the 2-field fermion-scalar extensions, some include a scalar singlet which could play the role of Dark Matter.
	
	An exercise in this direction was also performed in this work, in which we explored the full one-loop phenomenology of a 3-field extension which not only generates the bridge topology but can also explain the neutral $B$-anomalies, $R_K$ and $R_{K^{*}}$, and the CAA. This model, which we denote by the triple triplet, is constituted by a triplet leptoquark -- responsible for explaining the $B$-anomalies at tree-level --, a triplet vector-like lepton -- which accounts for the CAA at tree-level -- and a vector-like quark -- needed to generate the bridge topology. 
	
	Using {\tt matchmakereft} \cite{Carmona:2021xtq} to perform the full one-loop matching of this model -- which we provide in an auxiliary file -- and {\tt smelli} \cite{Aebischer_2019}, we have shown the allowed parameter region which not only explains the aforementioned anomalies but also respects bounds from an array of other observations.

	\section*{Acknowledgments}
	We are grateful to José Santiago, Javier Fuentes-Martín, Manuel Morales, Achilleas Lazopoulos, Mikael Chala, Susanne Westhoff and Maria Ramos for useful discussions.
	G.G is supported by LIP (FCT,
	COMPETE2020-Portugal2020, FEDER, POCI-01-0145-FEDER-007334) as well as by INCD under the project CPCA-A1-401197-2021, through Proyecto de Excelencia (P18-FR-4314) grant and by FCT
	under the project CERN/FIS-PAR/0032/2021 and under the grant SFRH/BD/144244/2019. P.O is supported by the Ministry of Science and Innovation and SRA (10.13039/501100011033) under grant PID2019-106087GB-C22, by the Junta de Andalucía grant FQM 101 and by an FPU grant from the Spanish government.
	
	\appendix
	\section{Box diagram results}
	\label{sec:boxes}

	\begin{figure}
		
		\centering
		\begin{subfigure}[b]{0.45\textwidth}
			\centering
			\includegraphics[width=\textwidth]{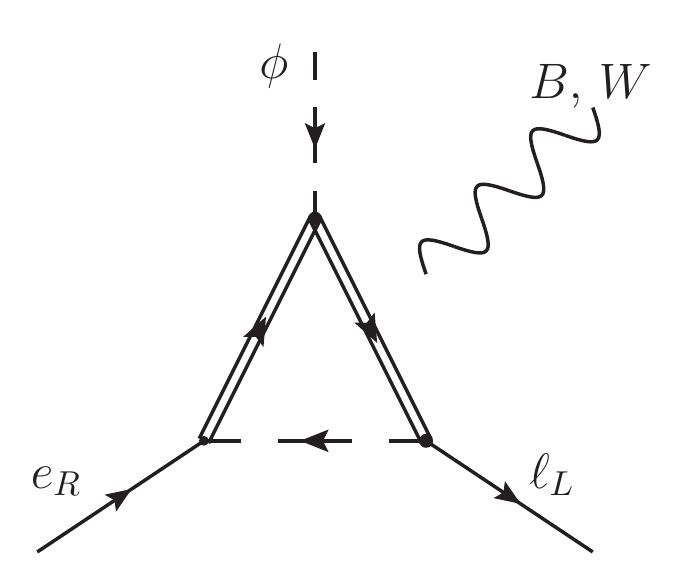}
			\caption{}
			\label{fig:triangleff}
		\end{subfigure}
		\begin{subfigure}[b]{0.45\textwidth}            
			\includegraphics[width=\textwidth]{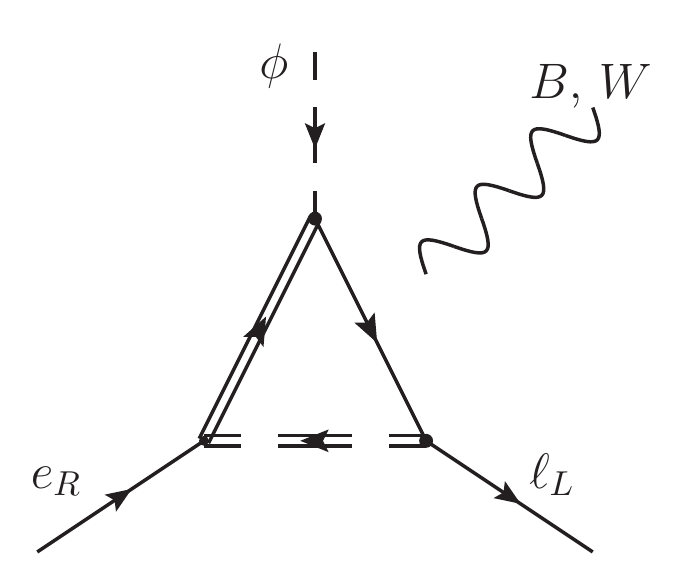}
			\caption{}
			\label{fig:trianglesf}
		\end{subfigure}%
		\caption{\emph{Left:} Box diagram contribution to $\alpha_{e\gamma}$ with 2 heavy fermion propagators. \emph{Right:} Box diagram contribution to $\alpha_{e\gamma}$ with 1 heavy fermion propagator and 1 heavy scalar propagator.  Double lines represent heavy particles whereas single lines are SM particles. The gauge boson ($B$ or $W$) is represented outside the diagram since it can be attached to any of the internal propagators. }\label{fig:bridgescalarw2fermionloop}
	\end{figure}
	
	\begin{figure}
		\centering
		\includegraphics[width=0.5\textwidth]{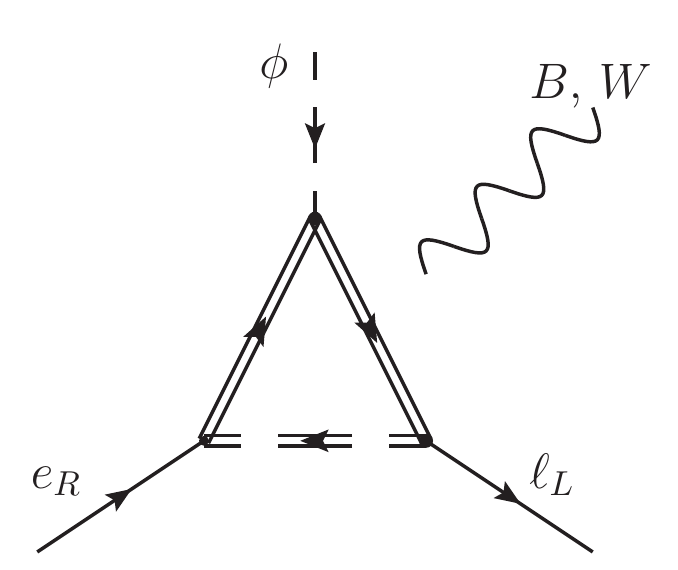}
		\label{fig:tfs}
		
		\caption{Box diagram contribution to $\alpha_{e\gamma}$ with all heavy internal propagators. Double lines represent heavy particles, whereas single lines are SM particles. The gauge boson ($B$ or $W$) is represented outside the diagram since it can be attached to any of the internal propagators.}\label{fig:trianglesff}
	\end{figure}
	For completeness, we present in this appendix the generic results for the contribution to $a_\mu$ by the box diagrams. Note that, while in our study these results are important for only some SM gauge representations of the BSM fields, these contributions are the ones commonly considered in the literature.
	
	For the box diagram with two heavy fermion propagators, Fig. \ref{fig:triangleff}, the schematic Lagrangian reads
	\begin{align}
		\mathcal{L} \supset&\, y_R T_{IJ} \overline{\Psi_1}_I \phi_J e_R  +  y_L T_{IJK} \overline{\ell_L}_{,I} \phi^\dagger_J {\Psi_2}_K + y_{H}^R T^H_{IJK} \overline{\Psi_2}_I \phi_J P_R {\Psi_1}_K \nonumber \\
		& + T^H_{IJK} y_{H}^L \overline{\Psi_2}_I \phi_J P_L {\Psi_1}_K + \mathrm{h.c.} \,,
	\end{align}
	where we use the same conventions for the gauge interactions of $\Psi$ and $\Phi$ as in Eq. (\ref{eq:lagdoublet}).

The contribution to $\alpha_{e\gamma}$ is given by:

	\begin{align}
		\alpha_{e\gamma}^{2,2}=& \left(\frac{i}{4}\right) ey_R y_L \sum_{\chi=R,L} y_H^\chi\left[T_{IJ}T_{2JK}T^\gamma_{I'I}T^H_{K2I'} \gamma_{\Psi_1}^\chi \right.\nonumber \\
		&\left. + T_{IJ}T_{2JK}T^H_{K'2I}T^\gamma_{KK'} \gamma_{\Psi_2}^\chi + T_{IJ}T_{2J'K}T^H_{K2I}T^\gamma_{J'J} \gamma_{\phi}^\chi \right], 
	\end{align}
	
	where $\chi$ sums over the right- and left-handed chiralities and the kinematic factors read:
	\begin{align}
		&\gamma_{\Psi_1}^L= 0 \,,\nonumber \\
		&\gamma_{\Psi_1}^R = -\frac{i}{16\pi^2} \frac{M_{\Psi_2}  \left(M_{\Psi_1}^2 \llog
			\left(\frac{M_{\Psi_1}^2}{M_{\Psi_2}^2}\right)-M_{\Psi_1}^2+M_{\Psi_2}^2\right)}{M_{\Psi_1}
			\left(M_{\Psi_1}^2-M_{\Psi_2}^2\right)^2} \,, \nonumber \\
		&\gamma_{\Psi_2}^L= 0 \,,\nonumber \\
		&\gamma_{\Psi_2}^R = -\frac{i}{16\pi^2} \frac{M_{\Psi_1} \left(-M_{\Psi_2}^2 \llog
			\left(\frac{M_{\Psi_1}^2}{M_{\Psi_2}^2}\right)+M_{\Psi_1}^2-M_{\Psi_2}^2\right)}{M_{\Psi_2}
			\left(M_{\Psi_1}^2-M_{\Psi_2}^2\right)^2} \, , \nonumber \\
		&\gamma_{\Phi}^L= 0 \,,\nonumber \\
		&\gamma_{\Phi}^R= \frac{-i}{16\pi^2 M_{\Psi_1}M_{\Psi_2} } \,.
	\end{align}
	
	For the box diagram with a light fermion in which the heavy fermion couples with the right-handed muon, the Lagrangian reads:
	\begin{align}
		\mathcal{L} \supset& y_R T^e_{IJ} \overline{\Psi_1}_I \Phi_J e_R +  y_L T^1_{IJ} \overline{\psi} \phi_J P_L {\Psi_1}_I + y_{\Phi} T^\Phi_{IJ} \overline{\ell_{LI}} \Phi_J^{\dagger} P_R \psi + \mathrm{h.c.} \,,
	\end{align}
	where $\psi$ is any light SM fermion which fits with the heavy field representations. The resulting contribution to $\alpha_{e\gamma}$ is:
	\begin{align}
		\alpha_{e\gamma}^{22}=& \left(\frac{i}{4}\right) e N y_R y_L y_\Phi  \left[T^e_{IJ} T^\gamma_{I'I} T^1_{I'2} T^\Phi_{2J} \gamma_{\Psi} \right.\nonumber \\
		&\left. + T^e_{IJ} T^1_{I2} Y_\psi T^\Phi_{2J} \gamma_{\psi} + T^e_{IJ} T^1_{I2}  T^\gamma_{JJ'} T^\Phi_{2J'}\gamma_{\Phi} \right], 
	\end{align}
	with the following kinematic factors:
	
	\begin{align}
		&\gamma_\Psi= -\frac{M_{\Phi}^2 \left(\left(M_{\Psi}^2+M_{\Phi}^2\right) \llog
			\left(\frac{M_{\Psi}^2}{M_{\Phi}^2}\right)-2 M_{\Psi}^2+2
			M_{\Phi}^2\right)}{\left(M_{\Phi}^2-M_{\Psi}^2\right)^3} \,,\nonumber \\
		&\gamma_\psi = \frac{-M_{\Phi}^2 \llog
			\left(\frac{M_{\Psi}^2}{M_{\Phi}^2}\right)+M_{\Psi}^2-M_{\Phi}^2}{\left(M_{\Psi}^2-M_{\Phi}^2\right)^2} \, , \nonumber \\
		&\gamma_\Phi =  \frac{M_{\Psi}^4-2 M_{\Psi}^2 M_{\Phi}^2 \llog
			\left(\frac{M_{\Psi}^2}{M_{\Phi}^2}\right)-M_{\Phi}^4}{\left(M_{\Psi}^2-M_{\Phi}^2\right)^3} \,.
	\end{align}

	In the case the heavy fermion couples with the left-handed muon, the relevant Lagrangian can be written as:
	\begin{align}
		\mathcal{L} \supset& \, y_R T^\ell_{IJK} \overline{\ell_L}_I \Phi_J^\dagger \Psi_{1K} +  y_L T^2_{IJK} \overline{\Psi}_I \phi_J P_L \psi_K   + y_{\Phi} T^\Phi_{IJ} \overline{\psi}_I \Phi_J e_R  + \mathrm{h.c.} \,,
	\end{align}
	resulting in the following contribution to $\alpha_{e\gamma}$:
	\begin{align}
		\alpha_{e\gamma}^{22}=& \left(\frac{i}{4}\right) e N y_R y_L y_\Phi  \left[T^\ell_{2JI} T^\gamma_{II'} T^2_{I'2K} T^\Phi_{KJ} \gamma_{\Psi} \right.\nonumber \\
		&\left. + T^\ell_{2JI} T^2_{I2K} T^\gamma_{KK'} T^\Phi_{K'J} \gamma_{\psi} + T^\ell_{2JI} T^2_{I2K} T^\Phi_{KJ'} T^\gamma_{J'J} \gamma_{\Phi} \right], 
	\end{align}
	where
	\begin{align}
		&\gamma_\Psi=  \frac{M_{\Phi}^2  \left(\left(M_{\Psi_1}^2+M_{\Phi}^2\right) \llog
			\left(\frac{M_{\Psi_1}^2}{M_{\Phi}^2}\right)-2 M_{\Psi_1}^2+2
			M_{\Phi}^2\right)}{\left(M_{\Psi_1}^2-M_{\Phi}^2\right)^3} \,,\nonumber \\
		&\gamma_\psi = -\frac{\left(M_{\Psi_1}^2 \llog
			\left(\frac{M_{\Psi_1}^2}{M_{\Phi}^2}\right)-M_{\Psi_1}^2+M_{\Phi}^2\right)}{\left(M_{\Psi_1}^2-M_{\Phi}^2\right)^2} \, , \nonumber \\
		&\gamma_\Phi =  \frac{M_{\Psi}^4-2 M_{\Psi}^2 M_{\Phi}^2 \llog
			\left(\frac{M_{\Psi}^2}{M_{\Phi}^2}\right)-M_{\Phi}^4}{\left(M_{\Psi}^2-M_{\Phi}^2\right)^3} \,.
	\end{align}

	When there are only heavy propagators in the box diagram, Fig. \ref{fig:trianglesff}, the relevant Lagrangian reads:
	\begin{align}
		\mathcal{L} &\supset y_R T^1_{IJ}\overline{\Psi_1}_I \Phi_J e_R +  y_L T^2_{IJ}\overline{\ell_L}\Psi_{2I} \Phi^\dagger_J + y^R_{H} T_{IJK}\overline{\Psi_2}_I \phi_J P_R {\Psi_1}_K\nonumber \\
		& + y^L_{H} T_{IJK} \overline{\Psi_2}_I \phi_J P_L {\Psi_1}_K + \mathrm{h.c.},
	\end{align}
	and the resulting $\alpha_{e\gamma}$ is given by:
	\begin{align}
		\alpha_{e\gamma}^{22}=& \left(\frac{i}{4}\right) y_R y_L \sum_{\chi=R,L} y_H^\chi\left[T^2_{IJ} T^H_{I2K} T^\gamma_{KK'} T^1_{K'J}  \gamma_{\Psi_1}^\chi \right.\nonumber \\
		&\left. + T^2_{IJ} T^\gamma_{II'} T_{I'2K}  T^1_{KJ} \gamma_{\Psi_2}^\chi + T^2_{IJ} T_{I2K} T_{KJ'2} T^\gamma_{JJ'} \gamma_{\Phi}^\chi \right], 
	\end{align}
	where
	
	\begin{align}
		\gamma_{\Psi_1}^L = &\frac{i}{16\pi^2} M_{\Phi}^2  \bigg[(M_{\Psi_2}-M_{\Psi_1}) (M_{\Psi_1}+M_{\Psi_2}) \left(M_{\Phi}^2
		(M_{\Psi_2}-M_{\Psi_1}) (M_{\Psi_1}+M_{\Psi_2}) \left(M_{\Psi_1}^2 \left(M_{\Phi}^2-2 
		M_{\Psi_2}^2\right) \right. \right. \bigg. \nonumber \\
		&\bigg. \left. \left.+M_{\Phi}^4\right) \llog \left(\frac{M_{\Psi_1}^2}{M_{\Phi}^2}\right)-(M_{\Phi}-M_{\Psi_1})
		(M_{\Psi_1}+M_{\Phi}) (M_{\Psi_2}-M_{\Phi}) (M_{\Psi_2}+M_{\Phi}) \left(M_{\Psi_1}^2 \left(M_{\Psi_2}^2-2
		M_{\Phi}^2\right) \bigg. \right. \right. \nonumber \\
		&\bigg. \left. \left. +M_{\Psi_2}^2 M_{\Phi}^2\right)\right)+M_{\Psi_2}^4 \left(M_{\Psi_1}^2-M_{\Phi}^2\right)^3
		\llog \left(\frac{M_{\Psi_1}^2}{M_{\Psi_2}^2}\right)\bigg]\times\nonumber\\
		&\frac{1}{(M_{\Psi_1}-M_{\Psi_2})^2 (M_{\Psi_1}+M_{\Psi_2})^2 \left(M_\Phi^2-M_{\Psi_1}^2\right)^3
			(M_{\Psi_2}-M_{\Phi})^2 (M_{\Psi_2}+M_{\Phi})^2}\,,
	\end{align}
	
	\begin{align}
		\gamma^R_{\Psi_1}= & \frac{i}{16\pi^2}M_{\Psi_1}  \left( M_{\Psi_2}(M_{\Psi_2}-M_{\Psi_1}) (M_{\Psi_1}+M_{\Psi_2})
		\left((M_{\Psi_1}-M_{\Phi}) (M_{\Psi_1}+M_{\Phi}) \right. \right.\nonumber \\
		&\left. \left. (M_{\Psi_2}-M_{\Phi}) (M_{\Psi_2}+M_{\Phi})
		\left(M_{\Psi_1}^2 M_{\Psi_2}^2-3 M_{\Psi_2}^2 M_{\Phi}^2+2 M_{\Phi}^4\right)+M_{\Phi}^4
		(M_{\Psi_2}-M_{\Psi_1}) (M_{\Psi_1}+M_{\Psi_2} ) \right. \right.  \nonumber \\
		& \left. \left. \left(M_{\Psi_1}^2+2 M_{\Psi_2}^2-3 M_{\Phi}^2\right) \llog
		\left(\frac{M_{\Psi_1}^2}{M_{\Phi}^2}\right)\right)+M_{\Psi_2}^3 \left(M_{\Psi_1}^2-M_{\Phi}^2\right)^3
		\left(M_{\Psi_2}^2-2 M_{\Phi}^2\right) \llog \left(\frac{M_{\Psi_1}^2}{M_{\Psi_2}^2}\right)\right) \times \nonumber \\
		& \frac{1}{(M_{\Psi_1}-M_{\Psi_2})^2 (M_{\Psi_1}+M_{\Psi_2})^2 (M_{\Psi_1}-M_{\Phi})^3 (M_{\Psi_1}+M_{\Phi})^3
			(M_{\Psi_2}-M_{\Phi})^2 (M_{\Psi_2}+M_{\Phi})^2}\,,
	\end{align}
	
	\begin{align}
		&\gamma^L_{\Psi_2} = -\frac{i}{16\pi^2} \left[M_{\Psi_1}^6 \left(-\left(M_{\Psi_2}^2-M_{\Phi}^2\right)^3\right) \llog
		\left(\frac{M_{\Psi_1}^2}{M_{\Phi}^2}\right)+M_{\Psi_1}^4 (M_{\Psi_1}-M_{\Phi}) (M_{\Psi_1}+M_{\Phi})
		\right. \nonumber \\
		& \left.  (M_{\Psi_2}-M_{\Phi})^3 (M_{\Psi_2}+M_{\Phi})^3 \llog
		\left(\frac{M_{\Psi_1}^2}{M_{\Psi_2}^2}\right)+(M_{\Psi_1}-M_{\Phi}) (M_{\Psi_1}+M_{\Phi}) \left(M_{\Phi}^2
		(M_{\Psi_1}-M_{\Psi_2}) (M_{\Psi_1}+M_{\Psi_2}) \right. \right .\nonumber \\
		&\left. \left. (M_{\Psi_2}-M_{\Phi}) (M_{\Psi_2}+M_{\Phi}) \left(M_{\Psi_1}^2
		\left(M_{\Psi_2}^2+M_{\Phi}^2\right)-2 M_{\Psi_2}^2 M_{\Phi}^2\right)+\left(M_{\Psi_1}^4
		M_{\Psi_2}^6-M_{\Phi}^6 \left(M_{\Psi_1}^2-M_{\Psi_2}^2\right)^2 \right. \right. \right.\nonumber \\
		& \left. \left. \left. +M_{\Psi_1}^2 M_{\Psi_2}^4 M_{\Phi}^2
		\left(M_{\Psi_2}^2-3 M_{\Psi_1}^2\right)+M_{\Psi_2}^2 M_{\Phi}^4 \left(M_{\Psi_1}^4+M_{\Psi_1}^2
		M_{\Psi_2}^2-M_{\Psi_2}^4\right)\right) \llog \left(\frac{M_{\Psi_2}^2}{M_{\Phi}^2}\right)\right) \right]\times \nonumber \\
		& \frac{1}{(M_{\Psi_1}-M_{\Psi_2})^2 (M_{\Psi_1}+M_{\Psi_2})^2 (M_{\Psi_1}-M_\Phi)^2 (M_{\Psi_1}+M_{\Phi})^2
			(M_{\Psi_2}-M_{\Phi})^3 (M_{\Psi_2}+M_{\Phi})^3}\,,
	\end{align}
	
	\begin{align}
		&\gamma^R_{\Psi_2} =\frac{i}{16\pi^2}M_{\Psi_1} M_{\Psi_2} \left(2 M_{\Psi_1}^2 (M_{\Psi_1}-M_{\Phi}) (M_{\Psi_1}+M_{\Phi})
		\left(M_{\Phi}^2-M_{\Psi_2}^2\right)^3 \llog \left(\frac{M_{\Psi_1}^2}{M_{\Psi_2}^2}\right) \right. \nonumber \\
		&\left. +M_{\Psi_1}^4
		\left(M_{\Psi_2}^2-M_{\Phi}^2\right)^3 \llog
		\left(\frac{M_{\Psi_1}^2}{M_{\Phi}^2}\right)+(M_{\Phi}-M_{\Psi_1}) (M_{\Psi_1}+M_{\Phi})
		\left((M_{\Psi_2}-M_{\Psi_1}) (M_{\Psi_1}+M_{\Psi_2}) \right. \right. \nonumber \\
		&\left. \left. (M_{\Psi_2}-M_{\Phi}) (M_{\Psi_2}+M_{\Phi})
		\left(M_{\Psi_1}^2 \left(M_{\Psi_2}^2-3 M_{\Phi}^2\right)+2 M_{\Phi}^4\right)+\left(M_{\Psi_1}^2
		M_{\Psi_2}^6+M_{\Psi_2}^4 M_{\Phi}^2 \left(M_{\Psi_2}^2-3 M_{\Psi_1}^2\right) \right. \right. \right. \nonumber \\
		& \left. \left. \left. +M_{\Phi}^4 \left(-2
		M_{\Psi_1}^4+6 M_{\Psi_1}^2 M_{\Psi_2}^2-3 M_{\Psi_2}^4\right)\right) \llog
		\left(\frac{M_{\Psi_2}^2}{M_{\Phi}^2}\right)\right)\right) \times \nonumber \\
		&\frac{1}{(M_{\Psi_1 }-M_{\Psi_2})^2 (M_{\Psi_1 }+M_{\Psi_2})^2 (M_{\Psi_1 }-M_\Phi)^2 (M_{\Psi_1 }+M_{\Phi})^2
			(M_{\Psi_2}-M_{\Phi})^3 (M_{\Psi_2}+M_{\Phi})^3}\,,
	\end{align}
	
	\begin{align}
		&\gamma^L_{\Phi} =\frac{iM_{\Phi}^2 }{16\pi^2}\left[ \frac{1}{\left(M_{\Psi_2}^2-M_{\Phi}^2\right)^3}\left(2 \left(M_{\Psi_1}^4 M_{\Psi_2}^4+M_{\Phi}^6 \left(M_{\Psi_1}^2+M_{\Psi_2}^2\right)-3
		M_{\Psi_1}^2 M_{\Psi_2}^2 M_{\Phi}^4\right) \llog
		\left(\frac{M_{\Psi_2}^2}{M_{\Phi}^2}\right) \right. \right. \nonumber \\
		&\left. \left. -(M_{\Phi}-M_{\Psi_1}) (M_{\Psi_1}+M_{\Phi})
		(M_{\Psi_2}-M_{\Phi}) (M_{\Psi_2}+M_{\Phi}) \left(M_{\Phi}^2 \left(M_{\Psi_1}^2+M_{\Psi_2}^2\right)-3
		M_{\Psi_1}^2 M_{\Psi_2}^2+M_{\Phi}^4\right) \right. \right. \nonumber \\
		&\left. \left. -\frac{2 M_{\Psi_1}^4 \llog
			\left(\frac{M_{\Psi_1}^2}{M_{\Psi_2}^2}\right)}{M_{\Psi_1}^2-M_{\Psi_2}^2}\right) \right] \times \frac{1}{\left(M_{\Psi_1}-M_\Phi\right)^3\left(M_{\Psi_1}+M_\Phi\right)^3}\,,
	\end{align}

	\begin{align}
		&\gamma^R_{\Phi} =\frac{i }{16\pi^2} M_{\Psi_1} M_{\Psi_2} \left(2 M_{\Psi_1}^2 M_{\Phi}^2
		\left(M_{\Psi_2}^2-M_{\Phi}^2\right)^3 \llog
		\left(\frac{M_{\Psi_1}^2}{M_{\Psi_2}^2}\right)+(M_{\Psi_1}-M_{\Psi_2}) (M_{\Psi_1}+M_{\Psi_2})\right. \nonumber \\
		& \left.  \left((M_{\Psi_1}-M_{\Phi}) (M_{\Psi_1}+M_{\Phi}) (M_{\Psi_2}-M_{\Phi}) (M_{\Psi_2}+M_{\Phi})
		\left(M_{\Phi}^2 \left(M_{\Psi_1}^2+M_{\Psi_2}^2\right)+M_{\Psi_1}^2 M_{\Psi_2}^2-3 M_{\Phi}^4\right) \right. \right. \nonumber \\
		&\left. \left. -2
		M_{\Phi}^2 \left(M_{\Psi_1}^4 M_{\Psi_2}^2+M_{\Psi_1}^2 \left(M_{\Psi_2}^4-3 M_{\Psi_2}^2
		M_{\Phi}^2\right)+M_{\Phi}^6\right) \llog \left(\frac{M_{\Psi_2}^2}{M_{\Phi}^2}\right)\right)\right) \times \nonumber \\
		& \frac{1}{(-M_{\Psi_1}^2 + M_{\Psi_2}^2) (M_{\Psi_1}^2 - M_{\Phi}^2)^3 (M_{\Psi_2}^2 - M_{\Phi}^2)^3}\,.
	\end{align}
	
	We cross-checked this last result of the box diagram with only heavy internal propagators with Eq. (4.4) of Ref. \cite{Allwicher:2021jkr} in the limit of degenerate masses and found perfect agreement.

	\bibliographystyle{style} 
	\bibliography{refsbridges} 
	
\end{document}